\pdfoutput=1

\documentclass[final,10pt]{elsarticle}
\makeatletter
\def\ps@pprintTitle{%
 \let\@oddhead\@empty
 \let\@evenhead\@empty
 \def\@oddfoot{}%
 \let\@evenfoot\@oddfoot}
\makeatother
\usepackage{graphicx}
\usepackage{rotating}
\usepackage{epstopdf, epsfig}
\usepackage[utf8]{inputenc} 
\usepackage[T1]{fontenc}    
\usepackage{hyperref}       
\usepackage{url}            
\usepackage{amsfonts}       
\usepackage{amsthm}       
\usepackage{mathtools}       
\usepackage{nicefrac}       
\usepackage{microtype}      
\usepackage{graphics}
\usepackage{pgfplots}
\usepackage{pgfplotstable}
\usepgfplotslibrary{fillbetween}
\usepackage{tikz}
\usepackage{bm}
\usepackage{xfrac}
\usepackage{float}
\usepackage{color}
\usepackage{algorithm}
\usepackage[noend]{algpseudocode}
\usepackage{mathtools}
\usepackage{amssymb}
\usepackage{amsmath}
\usepackage{standalone}
\usepackage{fullpage}
\usepackage[keeplastbox]{flushend}
\usepackage{comment}

\definecolor{green}{RGB}{0,102,0}

\usepackage{bm}
\usepackage{amssymb}
\usepackage{mathtools}
\usepackage{nicefrac}
\usepackage{subfig}

\newcommand*{\range}{\mathrm{Ran}}
\newcommand*{\tran}{^{\mathsf{T}}}

\newcommand*{\identity}{\bm{I}}			
\newcommand*{\zero}{\bm{0}}			
\newcommand*{\LSPGNameDeep}{Deep LSPG}			
\newcommand*{\DCNameCap}{Deep Conservation}			

\newcommand{\RR}[1]{\mathbb{R}^{#1}} 					
\newcommand{\RRplus}{\mathbb{R}_{+}} 					
\newcommand{\scalingOp}{\bm{S}} 					
\newcommand{\nDim}{d} 					

\newcommand{\nsnap}{n_\text{snap}} 					
 					
\newcommand{\fvmesh}{\mathcal M}

\newcommand{\NN}{\mathbb{N}}
\newcommand{\natNo}{\NN}
\newcommand{\nat}[1]{\natNo(#1)}

\newcommand{\innat}[1]{\in\nat{#1}}

\newcommand{\dofhf}{N} 					
\newcommand{\nstate}{\dofhf} 					
\newcommand{\ndof}{\dofhf} 					

\newcommand{\dofrom}{p}					
\newcommand{\nstatered}{p}					
\newcommand{\nseq}{N_t}					
\newcommand{\ntime}{N_t}					
\newcommand{\ntrain}{n_{\text{train}}}
\newcommand{\ninput}{\nstate}
\newcommand{\nlatent}{\nstatered}

\newcommand{\nepoch}{n_{\text{epoch}}}

\newcommand*{\states}{\bm{x}}				
\newcommand*{\state}{\states}				
\newcommand*{\initstate}{\states^{0}}	
\newcommand*{\stateRef}{\state_\mathrm{ref}}	
\newcommand*{\xoned}{x}
\newcommand*{\xtwod}{\vec{\xoned}}

\newcommand{\solSymb}{x}

\newcommand{\stateEntry}[1]{x_{#1}}
\newcommand{\stateArg}[1]{\state^{#1}}
\newcommand{\stateInit}{\stateArg{0}}
\newcommand{\stateInitEntry}[1]{\solSymb^0_{#1}}

\newcommand*{\gstate}{\bm{w}} 			

\newcommand*{\aprxstate}{\tilde{\states}}

\newcommand*{\eqvar}{w}
\newcommand*{\conservedQuantity}[1]{\eqvar_{#1}}

\newcommand{\conservedQuantityInitNo}{\eqvar^0}
\newcommand{\conservedQuantityInit}[1]{\conservedQuantityInitNo_{#1}}

\newcommand*{\velo}{\bm{f}}

\newcommand*{\paramSymb}{{\mu}}			
\newcommand*{\param}{\bm{\paramSymb}}			
\newcommand*{\gparam}{\bm{\nu}}			
\newcommand*{\paramelem}[1]{\paramSymb_{#1}}

\newcommand*{\gtime}{\tau}

\newcommand*{\paramtrain}{\param_{\text{train}}}
\newcommand*{\paramtest}{\param_{\text{test}}}
\newcommand*{\difftime}[1]{\dot{#1}}

\newcommand*{\spatialspace}{\Omega}
\newcommand*{\paramspace}{\mathcal D}	
\newcommand*{\nparam}{{n_\paramSymb}}	
\newcommand*{\paramspacetrain}{\paramspace_{\text{train}}}

\newcommand{\volume}[1]{|\Omega_{#1}|}
\newcommand{\controlVolSize}[1]{|{\spatialspace_{#1}}|}
\newcommand{\controlDecompVolSize}[1]{|{\bar{\spatialspace}_{#1}}|}
\newcommand*{\res}{\bm{r}}				
\newcommand*{\timestep}{\Delta t}				

\newcommand*{\rdstate}{\hat{\states}}		
\newcommand*{\grdstate}{\hat{\bm{\xi}}}

\newcommand*{\manifold}{\mathcal S}

\newcommand*{\defeq}{\vcentcolon =}

\newcommand*{\snapshotMat}{\bm{\mathrm{X}}}

\newcommand*{\stateSnapshot}[1]{\bm{\mathrm{x}}^{#1}}				
\newcommand*{\snapshotMatArg}[1]{\snapshotMat(#1)}

\newcommand*{\ddstates}{\bar{\states}}

\newcommand*{\nddstate}{\bar{\nstate}}

\newcommand*{\conserveOp}{\bar{\bm C}}
\newcommand*{\conserveOpGlobal}{\conserveOp_{\text{1}}}
\newcommand*{\globalmapState}[2]{\mathcal I_{(#1,#2)}}
\newcommand*{\globalmapDDState}[2]{\bar{\mathcal I}_{(#1,#2)}}

\newcommand*{\NSpatialElem}{N_{\spatialspace}}
\newcommand*{\nControlVol}{\NSpatialElem}

\newcommand*{\spatialspaceElem}[1]{\spatialspace_{#1}}
\newcommand*{\spatialspaceDDElem}[1]{\bar{\spatialspace}_{#1}}

\newcommand{\controlVolArg}[1]{\spatialspace_{#1}}

\newcommand*{\indicator}{I}

\newcommand{\meshDecomp}{\bar\fvmesh}
\newcommand{\meshDecompGlobal}{{\bar\fvmesh}_\text{global}}

\newcommand{\decomp}[1]{\bar{#1}}
\newcommand{\solDecompSymb}{\decomp{\solSymb}}
\newcommand{\stateDecomp}{\decomp{\states}}

\newcommand{\stateDecompEntry}[1]{\solDecompSymb_{#1}}

\newcommand{\nSubdomains}{\ensuremath{{N_{\decomp\Omega}}}}

\newcommand{\subdomainSymb}{\decomp{\Omega}}
\newcommand{\subdomainArg}[1]{\subdomainSymb_{#1}}

\newcommand{\subdomainInterfaceArg}[1]{\decomp\Interface_{#1}}

\newcommand{\subdomainFaceSet}{\bar\faceSet}
\newcommand{\subdomainFaceSetArg}[1]{\subdomainFaceSet_{#1}}
\newcommand{\subdomainSize}[1]{|\subdomainSymb_{#1}|}

\newcommand*{\lagrangeMul}{\bm{\lambda}}

\newcommand*{\hybridTerm}{\rho}

\newcommand*{\penaltyTerm}{\rho}

\newcommand*{\rdbasisnl}{\bm{d}}			
\newcommand*{\jacrdbasisnl}{\bm{J}}				
\newcommand*{\searchDirNK}{{\bm{p}}^{n(k)}}				
\newcommand*{\rdstateNK}{{\rdstate}^{n(k)}}				
\newcommand*{\rdstateNKp}{{\rdstate}^{n(k+1)}}				
\newcommand*{\rdstateNZero}{{\rdstate}^{n(0)}}				
\newcommand*{\linesearchNK}{{\alpha}^{n(k)}}				

\newcommand*{\rdbasis}{\bm{\Phi}}

\newcommand*{\rdtestbasis}{\bm{\Psi}}

\newcommand*{\dstate}{\bm{\xi}}
\newcommand*{\drdstate}{\hat{\bm{\xi}}}

\newcommand*{\autoenc}{\bm{h}}			
\newcommand*{\encoder}{\autoenc_{\text{enc}}}
\newcommand*{\decoder}{\autoenc_{\text{dec}}}

\newcommand*{\gvec}{\bm{x}}				

\newcommand*{\nnparam}{\bm{\theta}}

\newcommand*{\encparam}{\nnparam_{\text{enc}}}
\newcommand*{\encparamOpt}{\nnparam_{\text{enc}}^\star}
\newcommand*{\decparam}{\nnparam_{\text{dec}}}
\newcommand*{\decparamOpt}{\nnparam_{\text{dec}}^\star}

\newcommand{\learningrate}{\eta}

\newcommand{\nConservation}{\ensuremath{n_w}}

\newcommand{\normalVec}{\bm{n}}
\newcommand{\normalVecArg}[1]{\normalVec_{#1}}
\newcommand{\normalVecDecompArg}[1]{\decomp{\normalVec}_{#1}}

\newcommand*{\fluxSymb}{g}
\newcommand*{\fluxVec}{\bm{\fluxSymb}}
\newcommand*{\fluxVecArg}[1]{\fluxVec_{#1}}

\newcommand{\fluxApproxVecArg}[1]{\bm{\fluxSymb}_{#1}^\mathrm{FV}}

\newcommand{\fluxApprox}[2]{\bm{\fluxSymb}^\mathrm{FV}}

\newcommand{\veloSymb}{f}
\newcommand{\veloFluxSymb}{\veloSymb^\fluxSymb}
\newcommand{\sourceSymb}{s}
\newcommand{\sourceNo}{\sourceSymb}
\newcommand{\sourceEntry}[1]{\sourceNo_{#1}}
\newcommand{\sourceApproxNo}{\sourceSymb^\mathrm{FV}}
\newcommand{\sourceApproxEntry}[1]{\sourceApproxNo_{#1}}

\newcommand*{\veloFlux}{\velo^\fluxSymb}
\newcommand{\veloFluxEntryArgs}[4]{\veloFluxSymb_{#1}(#2,#3;#4)}

\newcommand{\veloSource}{\bm{\veloSymb}^{\sourceSymb}}
\newcommand{\veloSourceSymb}{{\veloSymb}^{\sourceSymb}}
\newcommand{\veloSourceEntryArgs}[4]{\veloSourceSymb_{#1}(#2,#3;#4)}

\newcommand{\controlVolSet}{\mathcal K}

\newcommand*{\dVol}{\, \mathrm{d}\xtwod}
\newcommand*{\dInterface}{\, \mathrm{d}\vec s(\xtwod)}

\newcommand{\Interface}{\Gamma}
\newcommand{\InterfaceArg}[1]{\Interface_{#1}}

\newcommand*{\generalSubdomain}{\omega}

\newcommand*{\generalInterface}{\gamma}

\newcommand{\face}{e}
\newcommand{\faceArg}[1]{\face_{#1}}
\newcommand{\faceSet}{\mathcal E}
\newcommand{\faceSetArg}[1]{\faceSet_{#1}}

\newcommand{\nFaces}{N_\face}

\newcommand{\ndofDecomp}{\decomp{\ndof}}

\newcommand{\consMapNo}{\mathcal I}
\newcommand{\consMap}[2]{\consMapNo(#1,#2)}
\newcommand{\consMapDecompNo}{\decomp{\consMapNo}}
\newcommand{\consMapDecomp}[2]{\consMapDecompNo(#1,#2)}

\newcommand{\velocityDecompSymb}{\decomp f}
\newcommand{\velocityDecompSymbol}{\bm{\velocityDecompSymb}}

\newcommand{\velocityDecompFluxSymb}{\decomp \veloSymb^\fluxSymb}
\newcommand{\velocityDecompSourceSymb}{\decomp \veloSymb^\sourceSymb}

\newcommand{\velocityDecompFlux}{\bm{\velocityDecompSymb}^{\fluxSymb}}
\newcommand{\velocityDecompFluxEntryArgs}[4]{\velocityDecompFluxSymb_{#1}(#2,#3;#4)}

\newcommand{\velocityDecompSource}{\bm{\velocityDecompSymb}^{\sourceSymb}}
\newcommand{\velocityDecompSourceEntryArgs}[4]{\velocityDecompSourceSymb_{#1}(#2,#3;#4)}

\newcommand*{\Error}{\mathcal E}
\newcommand*{\error}{\varepsilon}
\newcommand*{\stateError}{\Error_{\states}}

\newcommand*{\stateErrorGlobal}{\Error_{\states,\text{global}}}

\newcommand*{\resErrorGlobal}{\Error_{\res,\text{global}}}
\newcommand*{\resErrorNGlobal}{\error_{\res,\text{global}}}

\newcommand*{\LagrangeScalar}{\sigma}

\theoremstyle{definition}

\newtheorem{theorem}{Theorem}[section]

\newtheorem{remark}[theorem]{Remark}

\begin{document}
\numberwithin{equation}{section}
\begin{frontmatter}
	\title{Deep Conservation: A Latent-Dynamics Model \\for Exact Satisfaction of Physical Conservation Laws}

\author[sandia]{Kookjin Lee\corref{sandiacor}}
\ead{koolee@sandia.gov}
\author[uw]{Kevin T.\ Carlberg}
\ead{ktcarlb@uw.edu}
\ead[url]{kevintcarlberg.net}

\address[sandia]{Sandia National Laboratories}
\cortext[sandiacor]{7011 East Ave, MS 9159, Livermore, CA 94550.}
\address[uw]{University of Washington}

\begin{abstract}
This work proposes an approach for latent-dynamics learning that exactly
	enforces physical conservation laws. The method comprises two steps. First,
	the method computes a low-dimensional embedding of the high-dimensional
	dynamical-system state using deep convolutional autoencoders. This defines a
	low-dimensional nonlinear manifold on which the state is subsequently
	enforced to evolve. Second, the method defines a latent-dynamics model that
	associates with the solution to a constrained optimization problem.
	Here, the objective function is defined as the sum of squares of
	conservation-law violations over control volumes within a finite-volume
	discretization of the problem; nonlinear equality constraints explicitly
	enforce conservation over prescribed subdomains of the problem. Under modest
	conditions, the
	resulting dynamics model
guarantees that the time-evolution of the latent state exactly satisfies conservation laws over the prescribed subdomains. 
\end{abstract}

\end{frontmatter}

\section{Introduction}

Learning a latent-dynamics model for complex real-world physical processes (e.g., fluid dynamics \cite{lee2020model,morton2018deep,wiewel2019latent}, deformable solid mechanics \cite{fulton2019latent})
comprises an important task in science and engineering,
as it provides a mechanism for modeling the dynamics of
physical systems and can provide a rapid simulation tool for time-critical
applications such as control and design optimization. Two main ingredients
are required to learn a latent-dynamics model: (1) an \textit{embedding},
which provides a mapping between high-dimensional dynamical-system state and
low-dimensional latent variables, and (2) a \textit{dynamics model}, which
prescribes the time evolution of the latent variables in the latent space. 

There are two primary classes of methods for learning a latent-dynamics model.
The first class comprises \textit{data-driven dynamics learning}, which aims
to learn both the embedding and the dynamics model in a purely data-driven
manner that requires only measurements of the state/velocity. As such, this
class of methods does not require \textit{a priori} knowledge of the system of ordinary differential equations (ODEs) governing the high-dimensional dynamical system.  These methods typically learn a nonlinear embedding (e.g., via autoencoders
\cite{lusch2018deep,morton2018deep,otto2019linearly,takeishi2017learning}),
and---inspired by Koopman operator theory---learn a dynamics model that is
constrained to be linear. In a control \cite{lesort2018state} or
reinforcement-learning context \cite{bohmer2015autonomous}, 
the embedding and dynamics models can be learned simultaneously from 
observations of the state; however, most such models restrict the dynamics to
be locally linear
\cite{banijamali2017robust,goroshin2015learning,karl2016deep,watter2015embed}.

The second class of methods corresponds to \textit{projection-based dynamics
learning}, which learns the embedding in a data-driven manner, but computes
the dynamics model via a projection process executed on the governing system
of ODEs, which must be known \textit{a priori}. As opposed to the data-driven
dynamics learning, projection-based methods almost always employ a linear
embedding, which is typically defined by principal component analysis (or
``proper orthogonal decomposition'' \cite{holmes2012turbulence}) performed on
measurements of the state. The projection process that produces the
latent-dynamics model requires substituting the linear embedding in the
governing ODEs and enforcing orthogonality of the resulting residual to a
low-dimensional linear subspace \cite{benner2015survey}, yielding a
(Petrov--) Galerkin projection formulation.s

Each approach suffers from particular shortcomings. Because data-driven
dynamics learning lacks explicit \textit{a priori} knowledge of the governing ODEs---and thus predicts latent dynamics separately from any computational-physics code---these methods risk severe violation of physical principles underpinning the dynamical system. On the other hand, projection-based dynamics-learning
methods heavily rely on linear embeddings and, thus, exhibit limited dimensionality reduction compared with what is achievable with nonlinear embeddings \cite{ohlberger2016reduced}. Recently, this limitation has been
resolved by employing a nonlinear embedding (learned by deep convolutional
autoenoders) and projecting the governing ODEs onto the resulting
low-dimensional nonlinear manifold \cite{lee2020model}. Another shortcoming of
many projection-based dynamics-learning methods is that the (Petrov--)Galerkin projection process that 
they employ does not preclude
the violation of important physical properties such as conservation. To mitigate this issue, a recent work has proposed a projection technique that explicitly enforces conservation over subdomains by adopting a constrained least-squares formulation to define the projection \cite{carlberg2018conservative}.

In this study, we consider problems characterized by physical conservation
laws 
such problems are
ubiquitous in science and engineering.\footnote{In physics, conservation laws state that certain physical quantities of an isolated physical system do not change over time. In fluid dynamics, for example, the Euler equations \cite{leveque2002finite} governing inviscid flow are a set of equations representing the conservation of mass, momentum, and energy of the fluid.}  For such problems, we propose Deep Conservation: a projection-based dynamics learning method that combines
the advantages of Refs.~\cite{lee2020model} and
\cite{carlberg2018conservative}, as the method (1) learns a nonlinear
embedding via deep convolutional autoencoders, and (2) defines a dynamics
model via projection process that explicitly enforces conservation over
subdomains. The method assumes explicit \textit{a priori} knowledge of the
ODEs governing the conservations laws in integral form, and an associated
finite-volume discretization. In contrast to existing methods for
latent-dynamics learning, this is the only method that both employs a
nonlinear embedding and computes nonlinear dynamics for the latent state in a
manner that guarantees the satisfaction of prescribed physical properties.

Relatedly, we also note that there are deep-learning-based approaches for enforcing conservations laws by (1) designing neural networks that can learn arbitrary conservation laws (hyperbolic conservation laws  \cite{raissi2019physics}, Hamiltonian dynamics \cite{greydanus2019hamiltonian,toth2019hamiltonian}, Lagrangian dynamics \cite{mojgani2017lagrangian,cranmer2020lagrangian}), or (2) designing a loss function or adding an extra neural network constraining linear conservations laws \cite{beucler2019achieving}. These approaches, however, approximate solutions in a (semi-) supervised-learning setting and the resulting approximation does not guarantee exact satisfaction of conservation laws. Instead, the proposed latent-dynamics model associates with the solution to an constrained residual minimization problem, and guarantees the exact satisfaction of conservation laws over the prescribed subdomains under modest conditions. 

An outline of the paper is as follows. Section \ref{sec:fom} describes the
full-order model, which corresponds to a finite-volume discretizations of parameterized systems of 
physical conservation laws. Section \ref{sec:autoencoder} describes a nonlinear embedding constructed via 
using deep convolutional autoencoders. Section \ref{sec:rom} describes a projection-based nonlinear latent-dynamics model, which exactly enforces conservation laws. Section \ref{sec:numexp} provides results of numerical experiments on a benchmark advection problem that illustrate the method's ability to drastically reduce the dimensionality while successfully enforcing physical conservation laws. Finally, Section \ref{sec:conc}, we draw some conclusions.

\section{Full-order model}\label{sec:fom}
\subsection{Physical conservation laws}
This work considers parameterized systems of \textit{physical conservation laws}. The governing equations in integral form correspond to
\begin{equation}\label{eq:PDE}
	\frac{d}{dt}\int_{\generalSubdomain}{\conservedQuantity{i}}(\xtwod,t;\param)\dVol +\int_{\generalInterface}\fluxVecArg{i}(\xtwod,t;\param)\cdot\normalVec(\xtwod)\dInterface = \int_{\generalSubdomain}{\sourceEntry{i}}(\xtwod,t;\param)\dVol,
\end{equation}
for $i\innat{\nConservation},\ \forall\generalSubdomain\subseteq\spatialspace$,
which is solved in time domain $t\in[0, T]$ given an initial condition denoted by  $\conservedQuantityInit{i}:\spatialspace\times\paramspace\rightarrow\RR{}$ such that $\conservedQuantity{i}(\xtwod,0;\param) =
\conservedQuantityInit{i}(\xtwod;\param)$, $i\in\nat{\nConservation}$, where  $\nat{a}\defeq\{1,\ldots,a\}$. Here, $\generalSubdomain$ denotes any subset of the spatial domain  $\spatialspace\subset\RR{\nDim}$ with $\nDim\leq 3$;  $\generalInterface\defeq\partial \generalSubdomain$ denotes the boundary of the subset $\generalSubdomain$, while $\Interface\defeq\partial \spatialspace$ denotes the boundary of the domain $\spatialspace$; $\dInterface$ denotes integration with respect to the boundary; and $\conservedQuantity{i}:\spatialspace\times[0, T]\times
\paramspace\rightarrow \RR{}$, $\fluxVecArg{i}:\spatialspace\times[0, T]\times \paramspace\rightarrow \RR{\nDim}$, and $\sourceEntry{i}:\spatialspace\times[0, T]\times \paramspace \rightarrow \RR{}$ denote the
$i$th conserved variable, the flux associated with $\conservedQuantity{i}$, and the source associated with $\conservedQuantity{i}$. The parameters $\param\in\paramspace$ characterize physical properties of the
governing equations, where $\paramspace \subset \RR{\nparam}$ denotes the parameter space. Finally, $\normalVec:\generalInterface\rightarrow\RR{\nDim}$
denotes the outward unit normal to $\generalSubdomain$. We emphasize that equations \eqref{eq:PDE} describe conservation of \textit{any} set of variables $\conservedQuantity{i}$, $i\in\nat{\nConservation}$, given their respective flux $\fluxVecArg{i}$ and source $\sourceEntry{i}$ functions.

\subsection{Finite-volume discretization}\label{sec:fv_disc}
\begin{figure}[h]
    \centering
        \subfloat[1D spatial domain]  {\includegraphics[width=.35 \linewidth]{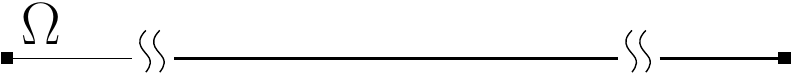}}
        \hspace{1cm}
    	\subfloat[Finite-volume mesh]  {\includegraphics[width=.35 \linewidth]{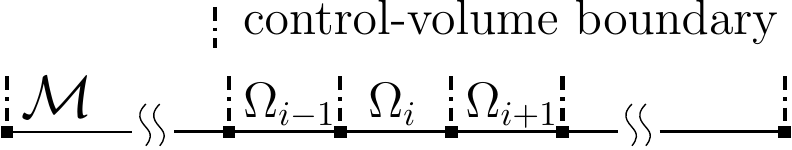}}
	\caption{An example one-dimensional spatial domain $\spatialspace$ and an example finite-volume mesh $\fvmesh = \{\controlVolArg{i}\}_{i=1}^{\NSpatialElem}$.}
	\label{fig:fvmesh}
\end{figure}

To discretize the governing equations \eqref{eq:PDE}, we apply the
finite-volume method \cite{leveque2002finite}, as it enforces conservation
numerically by decomposing the spatial
domain into many control volumes, numerically approximating the sources and
fluxes, and then enforcing conservation (Eq.\
\eqref{eq:PDE}) over the control volumes using the approximated quantities. In particular, we assume that the spatial domain $\spatialspace$ has been partitioned into a mesh $\fvmesh$ of $\NSpatialElem\in\natNo$ non-overlapping (closed, connected) control volumes $\controlVolArg{i}\subseteq\spatialspace$, $i\innat{\NSpatialElem}$. We define the mesh as
$\fvmesh \defeq \{\controlVolArg{i}\}_{i=1}^{\NSpatialElem}$,
and denote the boundary of the $i$th control volume by $\InterfaceArg{i}\defeq\partial\controlVolArg{i}$. 
The $i$th control-volume boundary is partitioned into a set of faces denoted by $\faceSetArg{i}$ such that $\InterfaceArg{i} =\{\xtwod\, |\, \xtwod\in \face,\ \forall \face\in\faceSetArg{i},\ i\innat{|\faceSetArg{i}|}\}$. Then the full set of $\nFaces$ faces within the mesh is $\faceSet\equiv\{\faceArg{i}\}_{i=1}^{\nFaces}\defeq \cup_{i=1}^{\NSpatialElem}\faceSetArg{i}$. Figure \ref{fig:fvmesh} depicts a one-dimensional spatial domain and a finite-volume mesh. 

Enforcing conservation \eqref{eq:PDE} on each control volume  yields
\begin{equation}\label{eq:PDEaftermesh}
    \frac{d}{dt}\int_{\controlVolArg{j}}{\conservedQuantity{i}}(\xtwod,t;\param)\dVol +\int_{\InterfaceArg{j}}\fluxVecArg{i}(\xtwod,t;\param)\cdot\normalVecArg{j}(\xtwod)\dInterface =\int_{\controlVolArg{j}}{\sourceEntry{i}}(\xtwod,t;\param)\dVol,
\end{equation}
for $i\innat\nConservation,\ j\innat\NSpatialElem$, where $\normalVecArg{j}:\InterfaceArg{j}\rightarrow\RR{\nDim}$ denotes the unit normal to control volume $\controlVolArg{j}$. Finite-volume schemes complete the discretization by  forming a  state vector $\state:[0, T]\times \paramspace\rightarrow
 \RR{\ndof}$ with $\ndof = \NSpatialElem\nConservation$ such that
\begin{equation}\label{eq:stateDefinition}
    \stateEntry{\consMap{i}{j}}(t;\param) =
    \frac{1}{\controlVolSize{j}}\int_{\controlVolArg{j}}{\conservedQuantity{i}}(\xtwod,t;\param)\dVol,
\end{equation}
where $\consMapNo:\nat{\nConservation}\times \nat{\NSpatialElem}\rightarrow\nat{\ndof}$ denotes a mapping from conservation-law index and control-volume index to degree of freedom, and a velocity vector $\velo:(\dstate,\gtime;\gparam)\mapsto\veloFlux(\dstate,\gtime;\gparam) + \veloSource(\dstate,\gtime;\gparam)$ with $\veloFlux,\veloSource:\RR{\ndof}\times[0,T]\times \paramspace \rightarrow  \RR{\ndof}$ whose elements consist of
\begin{align*}
	\begin{split}
	\veloFluxEntryArgs{\consMap{i}{j}}{\state}{t}{\param}&= -\frac{1}{\controlVolSize{j}}\int_{\InterfaceArg{j}}\fluxApproxVecArg{i}(\state;\xtwod,t;\param)\cdot\normalVecArg{j}(\xtwod)\dInterface, \\
\veloSourceEntryArgs{\consMap{i}{j}}{\state}{t}{\param}&=
\frac{1}{\controlVolSize{j}}\int_{\controlVolArg{j}}{\sourceApproxEntry{i}}(\state;\xtwod,t;\param)\dVol,
	\end{split}
\end{align*}
 for $i\innat\nConservation,\ j\innat\NSpatialElem$.  Here, $\fluxApproxVecArg{i}:\RR{\ndof}\times\spatialspace\times[0,T]\times \paramspace\rightarrow \RR{\nDim}$ and $\sourceApproxEntry{i}:\RR{\ndof}\times\spatialspace\times[0,T]\times \paramspace\rightarrow
\RR{}$, $i\innat{\nConservation}$ denote the approximated flux and source, respectively, associated with the $i$th conserved variable.

Substituting $\int_{\controlVolArg{j}}{\conservedQuantity{i}}(\xtwod,t;\param)\dVol\leftarrow\controlVolSize{j}\stateEntry{\consMap{i}{j}}(t;\param)$, $\fluxVecArg{i}\leftarrow \fluxApproxVecArg{i}$, and $\sourceEntry{i}\leftarrow\sourceApproxEntry{i}$ in Eq.~\eqref{eq:PDEaftermesh} and dividing by $\controlVolSize{j}$ yields 
\begin{equation}\label{eq:govern_eq}
	\difftime{\states} = \velo(\states,t;\param), \qquad \states(0;\param) = \initstate(\param),
\end{equation}
where $\stateInitEntry{\consMap{i}{j}}(\param) \defeq\frac{1}{\volume{i}}\int_{\controlVolArg{j}}\conservedQuantityInit{i}(\xtwod;\param)\dVol$ denotes the parameterized initial condition. This is a parameterized system
of nonlinear ordinary differential equations (ODEs) characterizing an initial value problem, which is our full-order model (FOM). We refer to Eq.~\eqref{eq:govern_eq} as the FOM ODE.

Numerically solving the FOM ODE~\eqref{eq:govern_eq} requires application of a time-discretization method. For simplicity, this work restricts attention to linear multistep methods. A linear $k$-step method applied to numerically solve the FOM ODE \eqref{eq:govern_eq} leads to solving the system of algebraic
equations
\begin{align}\label{eq:fomODeltaE}
    \res^{n}(\state^{n};\param) = \zero,\quad n=1,\ldots,\nseq,
\end{align}
where the time-discrete residual $\res^{n}: \RR{\nstate}\times\paramspace\rightarrow\RR{\nstate}$, as a function of $\dstate$ parameterized by $\gparam$, is defined as
\begin{align}\label{eq:disc_res}
        \res^{n}:(\dstate;\gparam) \mapsto &\alpha_0 \dstate - \Delta t \beta_0
    	\velo(\dstate,t^n;\gparam) + \sum_{j=1}^{k} \alpha_j \states^{n-j} - \Delta t \sum_{j=1}^{k} \beta_j \velo (\states^{n-j}, t^{n-j}; \gparam).
\end{align}
Here, $\timestep \in \mathbb{R}_{+}$ denotes the time step, $\states^{k}$ denotes the numerical approximation to $\states(k\timestep; \param)$, and the coefficients $\alpha_j$ and $\beta_j$, $j=0,\ldots, k$ with $\sum_{j=0}^k\alpha_j=0$ define a particular linear multistep scheme. These methods are implicit if $\beta_0\neq 0$. We refer to Eq.~\eqref{eq:fomODeltaE} as the FOM O$\Delta$E.

\subsection{Computational barrier: time-critical problems}
Many problems in science and engineering are \textit{time critical} in nature, meaning that the solution to the FOM O$\Delta$E \eqref{eq:fomODeltaE} must be computed within a specified computational budget (e.g., less than $0.1$
core--hours) for arbitrary parameter instances $\param\in\paramspace$. When the full-order model is truly high
fidelity, the computational mesh $\fvmesh$ often becomes very fine, which can yield an extremely large state-space dimension $\nstate$ (e.g., $\nstate\sim 10^7$). This introduces a \textit{de facto} computational barrier: the full-order model is too computationally expensive to solve within the prescribed computational budget. Such cases demand a method for \textit{approximately} solving the full-order model while retaining high levels of accuracy.
 
\begin{figure}[!h]
	\centering
	\vspace{1cm}
	\includegraphics[width=.65 \linewidth]{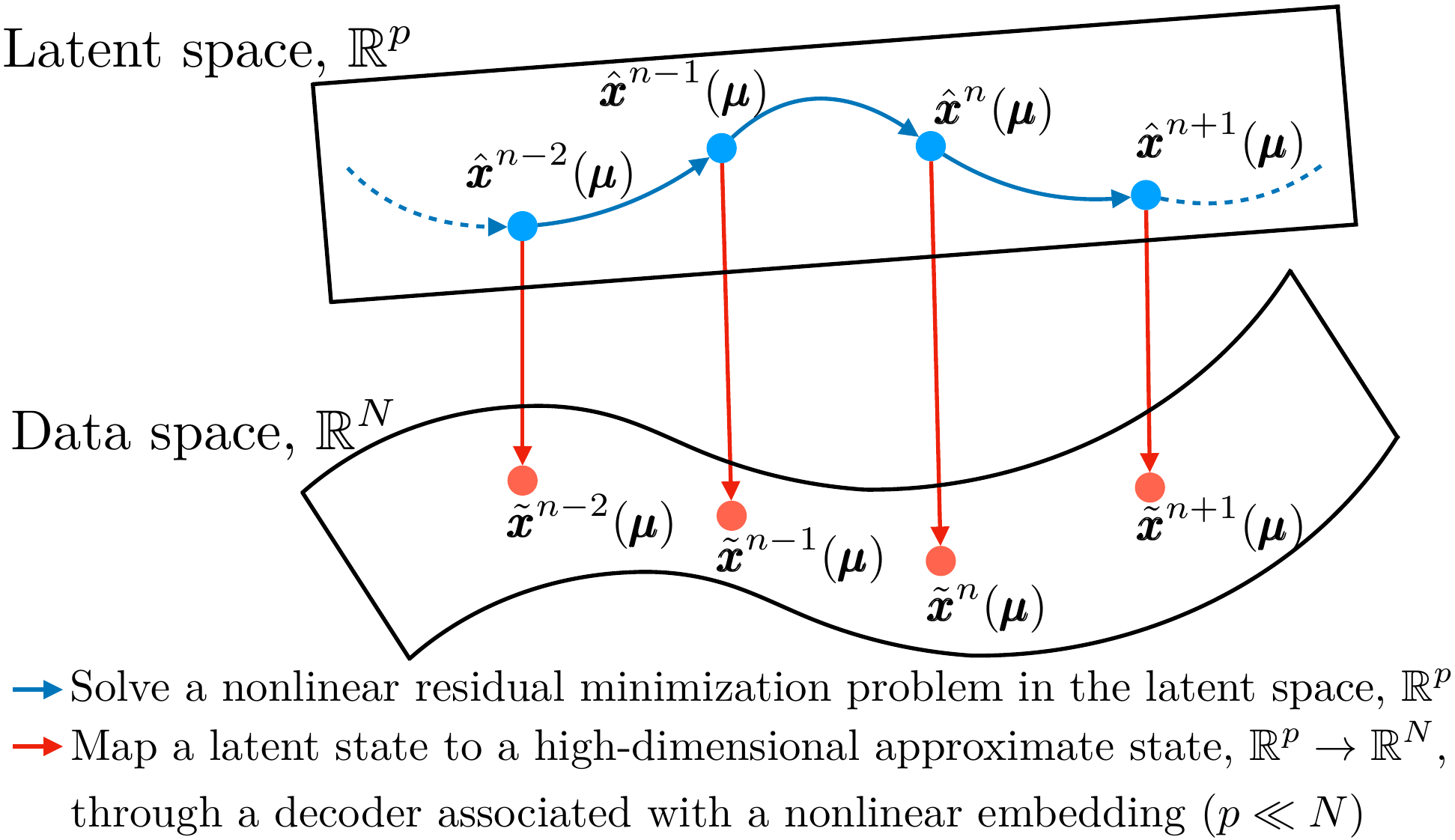}
	\vspace{1cm}
	\caption{Deep Conservation model -- the second stage: a latent dynamics (blue arrows) and a decoder (red arrows, the decoder associated with a nonlinear embedding). 
	}
	\label{fig:overview} 
\end{figure}

We now present a two-stage method that
(1) computes a nonlinear embedding of the state using deep convolutional autoencoders, and (2) computes a dynamics model for latent states that exactly satisfy the physical conservation laws over \textit{subdomains} comprising unions of control volumes of the mesh. Figure \ref{fig:overview} depicts the second stage of the proposed method, where the latent space is identified by convolutional autoencoders during the first stage; the method computes latent states $\{\rdstate^n(\param)\}_{n=1}^{\nseq}$ via conservation-enforcing projection (Section \ref{sec:deepcons}), and computes high-dimensional approximate states $\{\aprxstate^n(\param)\}_{n=1}^{\nseq}$ through a decoder associated with the nonlinear embedding (Section \ref{sec:nonlinear_embedding}).

\section{Nonlinear embedding: deep convolutional autoencoders}\label{sec:autoencoder}

\subsection{Deep convolutional autoencoders}
Autoencoders \cite{demers1993non,hinton2006reducing} consist of two parts: an encoder $\encoder(\cdot;\encparam):\RR{\ninput}\rightarrow\RR{\nlatent}$ and a decoder $\decoder(\cdot;\decparam):\RR{\nlatent}\rightarrow\RR{\ninput}$ with latent-state dimension $\nlatent\ll\ninput$ such that 
\begin{equation*}
    \autoenc:(\gvec;\encparam,\decparam) \mapsto \decoder(\cdot; \decparam) \circ \encoder(\gvec; \encparam),
\end{equation*}
where $\encparam$ and $\decparam$ denote parameters associated
with the encoder and decoder, respectively.

Because we are considering finite-volume discretizations of conservation laws, the state elements $\stateEntry{\consMap{i}{j}}$, $j\in\nat{\nControlVol}$ correspond to the value of the $i$th conserved variable $\conservedQuantity{i}$ distributed across the $\nControlVol$ control volumes characterizing the mesh $\fvmesh$. As such, we can interpret the state $\state\in\RR{\nstate}$ as representing the distribution of spatially distributed data with $\nConservation$ channels. This is precisely the format required by convolutional neural networks, which often generalize well to unseen test data \cite{lecun2015deep} because they exploit local connectivity, employ parameter sharing, and exhibit translation equivariance \cite{goodfellow2016deep,lecun2015deep}. Thus, we leverage the connection between conservation laws and image data, and employ convolutional autoencoders.

\subsection{Offline training}\label{sec:offline}
The first step of offline training is snapshot-based data collection. This requires solving the FOM O$\Delta$E \eqref{eq:fomODeltaE} for training-parameter instances $\param\in\paramspacetrain \equiv\{\paramtrain^{i}\}_{i=1}^{\ntrain} \subset \paramspace$ and assembling the snapshot matrix
\begin{equation} \label{eq:snapshotMat}
	\snapshotMat \defeq \left[\snapshotMatArg{\paramtrain^{1}}\ \cdots\
	\snapshotMatArg{\paramtrain^{\ntrain}}
	\right] \in \mathbb{R}^{\dofhf \times \nsnap}
\end{equation} 
with $\nsnap\defeq\nseq\ntrain$ and
$
	\snapshotMatArg{\param} 
	\equiv[\stateSnapshot{1}(\param) \ \cdots\
\stateSnapshot{\nseq}(\param)]
	\defeq [
	\state^{1}(\param) - \state^{0}(\param)\ 
	\cdots\ 
	\state^{\nseq}(\param) - \state^{0}(\param)] \in \RR{\dofhf \times \nseq}.
$		 

To improve numerical stability of the gradient-based optimization for
training, the first layer of the proposed autoencoder applies data
standardization through an affine
scaling operator $\scalingOp$, which ensures that all elements of the training data
lie between zero and one.  Then the
autoencoder reformats the input vector into a tensor compatible with
convolutional layers; the last layer applies the inverse scaling
operator $\scalingOp^{-1}$ and reformats the data into a vector.

Given the network architecture $\autoenc(\gvec;\encparam,\decparam)$, we
compute parameter values $(\encparamOpt,\decparamOpt)$ by  approximately solving
\begin{equation}\label{eq:AE_obj}
\underset{\encparam,\decparam}{\text{minimize}}\,\sum_{i=1}^{\nsnap}\|
\stateSnapshot{i} - \autoenc(\stateSnapshot{i};\encparam,\decparam)
\|_2^2
\end{equation}
using stochastic gradient descent (SGD) with minibatching and early stopping.

Along with this vanilla autoencoder, inspired by the formulation of  physics-informed neural networks \cite{raissi2019physics}, we devise another autoencoder with an additional training objective function, 
\begin{align*}
\underset{\encparam,\decparam}{\text{minimize}}\,&\sum_{i=1}^{\nsnap}\|
\stateSnapshot{i} - \autoenc(\stateSnapshot{i};\encparam,\decparam) 
\|_2^2 +\hybridTerm\sum_{i=1}^{\nsnap}\|\res^{i}(\scalingOp^{-1}(\autoenc(\stateSnapshot{i};\encparam,\decparam);\param^{i})) \|_2^2,
\end{align*}
which enforces minimization of the time-discrete residuals. The
advantage of this approach is that it aligns the training objective more
closely with the online objective (described in Section \ref{sec:deeplspg});
the disadvantage is that this approach is \textit{intrusive} as evaluating the objective function requires evaluating
the underlying finite-volume model. 
On the other hand, training the autoencoder with the original objective function is a purely data-oriented approach, which only requires solution snapshots and is agnostic to the finite-volume model and other problem specific information.

\subsection{Nonlinear embedding}\label{sec:nonlinear_embedding}
Given the trained autoencoder $\autoenc:(\gvec;\encparamOpt,\decparamOpt)\mapsto\decoder(\cdot;\decparamOpt) \circ \encoder(\gvec; \encparamOpt)$, we construct a nonlinear embedding by defining a low-dimensional nonlinear ``trial manifold'' on which we will restrict the
approximated state to evolve. In particular, we define this manifold from the extrinsic view as
$\manifold \defeq\{
\rdbasisnl(\grdstate)\,|\,
\grdstate\in\RR{\dofrom}\}$, where
the parameterization function is defined from the decoder as $\rdbasisnl:\grdstate\mapsto\decoder(\grdstate; \decparamOpt)$ with $\rdbasisnl:\RR{\nstatered}\rightarrow\RR{\nstate}$.
We subsequently approximate the state as $\state(t; \param)\approx\aprxstate(t; \param)\in \stateRef(\param)+ \manifold$, where $\stateRef(\param)=\stateInit(\param) -\encoder(\stateInit(\param); \encparamOpt)$ is the reference state. This approximation can be expressed algebraically as
\begin{equation}\label{eq:state_approx}
	\aprxstate(t; \param) = \stateRef(\param)+ \rdbasisnl(\rdstate(t;\param)),
\end{equation}
which elucidates the mapping between the latent state
$\rdstate:\RRplus \times \paramspace\rightarrow\RR{\dofrom}$
and the approximated state $\aprxstate:\RRplus\times\paramspace\rightarrow\RR{\nstate}$.

\begin{remark}[Linear embedding via POD]\label{rem:pod}
	Classical methods for projection-based dynamics learning employ proper orthogonal decomposition (POD) \cite{holmes2012turbulence}---which is closely related to principal component analysis---to construct a linear
	embedding. Using the above notation, POD computes the singular value decomposition of the snapshot matrix $\snapshotMat = \bm{U} \bm{\Sigma}\bm{V}$ and sets a ``trial basis matrix''  $\rdbasis\in\RR{\ndof\times\nstatered}$ 
	to be equal to the first $\nstatered$ columns of $\bm{U}$.
	Then, these methods define low-dimensional affine ``trial subspace'' such that the state is approximated as
	$\state(t; \param)\approx\aprxstate(t; \param)\in \stateInit(\param)+ \range{(\rdbasis)}$, which is equivalent to the approximation in Eq.~\eqref{eq:state_approx} with 
	$\stateRef(\param)=\stateInit(\param)$ and a linear parameterization function $\rdbasisnl:\drdstate\mapsto\rdbasis\drdstate$.
\end{remark}

\section{Latent-dynamics model: conservation-enforcing projection} \label{sec:rom}
We now describe the proposed projection-based dynamics model, starting with deep least-squares Petrov--Galerkin (LSPG) projection (proposed in Ref.~\cite{lee2020model}) in Section  \ref{sec:deeplspg}, and proceeding with the proposed Deep Conservation projection in Section \ref{sec:deepcons}.

\subsection{Deep LSPG projection}\label{sec:deeplspg}
To construct a latent-dynamics model for the approximated state 
$\aprxstate$, the Deep LSPG method \cite{lee2020model} simply substitutes $\states \leftarrow \aprxstate$ defined in Eq.~\eqref{eq:state_approx} into the FOM O$\Delta$E \eqref{eq:fomODeltaE} and minimizes the $\ell^2$-norm of the residual, i.e.,
\begin{equation}\label{eq:disc_opt_problem}
	\rdstate^n(\param)  =
	\underset{\grdstate\in\RR{\nstatered}}{\arg\min} \left\|
	\res^{n}\left(
	\stateRef(\param)+\rdbasisnl(\grdstate);\param\right)\right\|_2^2,
\end{equation}
which is solved sequentially for $n=1,\ldots,\nseq$. 

Eq.~\eqref{eq:disc_opt_problem} defines the (discrete-time) dynamics model for the latent states associated with Deep
LSPG projection. The nonlinear least-squares problem \eqref{eq:disc_opt_problem} can be solved
using, for example, the Gauss--Newton method, which leads to the iterations, for $k=0,\ldots,K$,
\begin{align*} 
\begin{split} 
	\rdtestbasis^{n}(\rdstateNK;\param)\tran
	\rdtestbasis^{n}(\rdstateNK;\param) \searchDirNK &=
	-
	\rdtestbasis^{n}(\rdstateNK;\param)\tran 
\res^{n}
	\left(\stateRef(\param)+\rdbasisnl(\rdstateNK);\param\right),\\
	\rdstateNKp &= \rdstateNK + \linesearchNK\searchDirNK.
\end{split} 
\end{align*} 

Here, $\rdstateNZero$ is the initial guess (often taken to be ${\rdstate}^{n-1}$); $\linesearchNK\in\RR{}$ is a step length chosen to satisfy the strong Wolfe conditions for global convergence; and $\rdtestbasis^{n}:\RR{\nstatered}\times\paramspace\rightarrow\RR{\nstate\times\nstatered}$, as a function of $\grdstate$ parameterized by $\gparam$, is 
\begin{align*}
	\rdtestbasis^{n}(\grdstate;\gparam) &=\frac{\partial \res^n}{\partial \dstate}(
	\stateRef(\gparam)+\rdbasisnl(\grdstate);\gparam)\jacrdbasisnl(\grdstate)\\
	&= \left( \alpha_0\identity - \Delta t \beta_0 \frac{\partial \velo}{\partial \dstate}
	\left( \stateRef(\gparam)+\rdbasisnl(\grdstate), t^n;\gparam \right) \right)\jacrdbasisnl (\grdstate),
\end{align*}
where $\jacrdbasisnl:\grdstate\mapsto \frac{d\rdbasisnl}{d
\drdstate}(\grdstate)$ is  the Jacobian of the decoder and $\dstate \in \RR{\dofhf}$.

\begin{remark}[POD--LSPG projection]
	POD--LSPG projection \cite{carlbergGalDiscOpt} employs the same residual-minimization projection \eqref{eq:disc_opt_problem}, but with 
	reference state $\stateRef(\param)=\stateInit(\param)$ and linear parameterization function $\rdbasisnl:\drdstate\mapsto \rdbasis\drdstate$ as described in Remark \ref{rem:pod}.
\end{remark}

\subsection{Deep Conservation projection}\label{sec:deepcons}

We now derive the proposed Deep Conservation projection, which effectively combines Deep LSPG projection \cite{lee2020model} just described with conservative LSPG projection \cite{carlberg2018conservative}, which was developed for linear embeddings only.

\begin{figure}[h]
	\centering
	\subfloat[Decomposed mesh $\meshDecomp$] 
	{\includegraphics[width=.35 \linewidth]{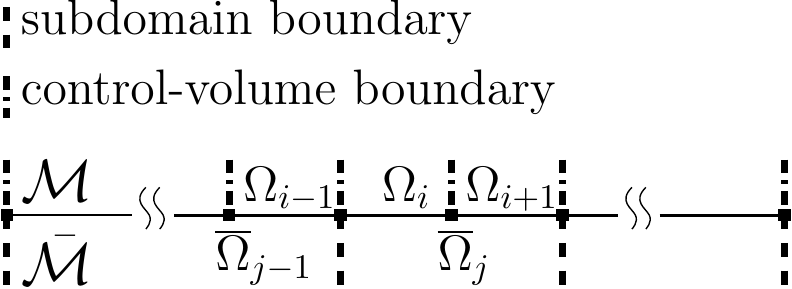}}
	\hspace{1cm}
	\subfloat[Decomposed mesh $\meshDecompGlobal$ with $\nSubdomains=1$, and $\subdomainArg{1}=\spatialspace$]  {\includegraphics[width=.35 \linewidth]{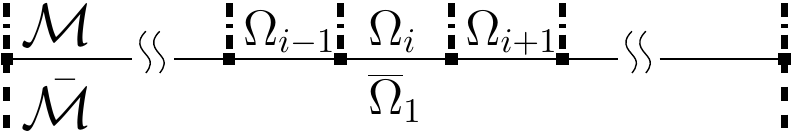}}
	\caption{Examples of decomposed meshes $\meshDecomp$ for the finite-volume mesh shown in Figure \ref{fig:fvmesh}.}
	\label{fig:ddmesh}
\end{figure}

To begin, we decompose the mesh $\fvmesh$ into subdomains, each of which comprises the union of control volumes. That is, we define a decomposed mesh $\meshDecomp$ of $\nSubdomains(\leq\nControlVol)$ subdomains $\subdomainArg{i}= \cup_{j\in\controlVolSet\subseteq\nat{\nControlVol}}\controlVolArg{j}$, $i\innat{\nSubdomains}$ with $\meshDecomp \defeq\{\subdomainArg{i}\}_{i=1}^{\nSubdomains}$. Denoting
the boundary of the $i$th subdomain by $\subdomainInterfaceArg{i}\defeq\partial\subdomainArg{i}$, we have $\subdomainInterfaceArg{i} = \{\xtwod\,|\, \xtwod \in \face,\ \forall \face\in\subdomainFaceSetArg{i},\ i\innat{|\subdomainFaceSetArg{i}|}\}\subseteq\cup_{j=1}^{\nControlVol} \InterfaceArg{j}$, $i\innat\nSubdomains$ with $\subdomainFaceSetArg{i}\subseteq\faceSet$ representing the set of faces belonging to the $i$th subdomain. We denote the full set of faces within the decomposed mesh by $\subdomainFaceSet\defeq\cup_{i=1}^{\nSubdomains}\subdomainFaceSetArg{i}\subseteq\faceSet$. Note that the global domain can be considered by employing $\meshDecomp = \meshDecompGlobal$, which is characterized by $\nSubdomains=1$ subdomain that corresponds to the global domain. Figure \ref{fig:ddmesh} depicts example decomposed meshes.

Enforcing conservation \eqref{eq:PDE} on each subdomain in the decomposed mesh yields, for $i\innat\nConservation,\ j\innat\nSubdomains$,
\begin{equation}\label{eq:PDEaftermeshDecomp}
    \frac{d}{dt}\int_{\subdomainArg{j}}{\conservedQuantity{i}}(\xtwod,t;\param)\dVol + \int_{\subdomainInterfaceArg{j}}\fluxVecArg{i}(\xtwod,t;\param)\cdot\normalVecDecompArg{j}(\xtwod)\dInterface =\int_{\subdomainArg{j}}{\sourceEntry{i}}(\xtwod,t;\param)\dVol,
\end{equation}
where $\normalVecDecompArg{j}:\subdomainInterfaceArg{j}\rightarrow\RR{\nDim}$ denotes the unit normal to subdomain $\subdomainArg{j}$. We propose applying the same finite-volume discretization employed to discretize the control-volume conservation equations \eqref{eq:PDEaftermesh} to the subdomain conservation equations \eqref{eq:PDEaftermeshDecomp}. To
accomplish this, we introduce a ``decomposed'' state vector
$\stateDecomp:\RR{\ndof}\times [0,T]\times \paramspace\rightarrow
 \RR{\ndofDecomp}$ with
 $\ndofDecomp = \nSubdomains\nConservation$ and
 elements, for $i\innat\nConservation,\ j\innat\nSubdomains$, 
\begin{equation}\label{eq:stateDecompDefinition}
    \stateDecompEntry{\consMapDecomp{i}{j}}(\states,t;\param) = \frac{1}{\subdomainSize{j}}\int_{\subdomainArg{j}}{\conservedQuantity{i}}(\xtwod,t;\param)\dVol, 
\end{equation}
where $\consMapDecompNo:\nat{\nConservation}\times\nat{\nSubdomains}\rightarrow\nat{\ndofDecomp}$ denotes a mapping from conservation-law index and subdomain index to decomposed degree of freedom. The decomposed state vector can be computed from the state vector $\states$ as
\begin{equation*}
	\ddstates(\states) = \conserveOp \states
\end{equation*}
where $\conserveOp \in \RR{\nddstate \times \nstate}_{+}$ has elements $\conserveOp_{\globalmapDDState{i}{j},\globalmapState{l}{k}} =  \frac{\vert\spatialspaceElem{k}\vert}{\vert\spatialspaceDDElem{j}\vert} \delta_{il}  \indicator(\spatialspaceElem{k} \subseteq \spatialspaceDDElem{j})$, where $\indicator$ is the indicator function, which outputs one if its argument is true, and zero otherwise. 

Similarly, the velocity vector $\velocityDecompSymbol(\states,t;\param)$  associated with the finite-volume scheme applied to  the decomposed mesh $\meshDecomp$, can be obtained by enforcing conservation \eqref{eq:PDE} on each subdomain as in Eq~\eqref{eq:PDEaftermeshDecomp} such that $\velocityDecompSymbol:
(\gstate,\gtime;\gparam)\mapsto\velocityDecompFlux(\gstate,\gtime;\gparam) +\velocityDecompSource(\gstate,\gtime;\gparam)$ with
$\velocityDecompFlux,\velocityDecompSource:\RR{\ndof}\times[0,T]\times\paramspace \rightarrow  \RR{\nddstate}$, whose elements consist of
\begin{align}\label{eq:velocityDefinitionC}
	\begin{split}
	\velocityDecompFluxEntryArgs{\consMap{i}{j}}{\state}{t}{\param}&=
		 -\frac{1}{\controlDecompVolSize{j}}\int_{\subdomainInterfaceArg{j}}\fluxApproxVecArg{i}(\state;\xtwod,t;\param)\cdot\normalVecArg{j}(\xtwod)\dInterface \\
	\velocityDecompSourceEntryArgs{\consMap{i}{j}}{\state}{t}{\param}&=
		\frac{1}{\subdomainSize{j}}\int_{\subdomainArg{j}}{\sourceApproxEntry{i}}(\state;\xtwod,t;\param)\dVol
	\end{split}
\end{align}
for $i\innat\nConservation,\ j\innat\nSubdomains$. Using the same matrix $\conserveOp \in \RR{\nddstate \times \nstate}_{+}$ in Section \ref{sec:deepcons}, $\velocityDecompFlux$ and $\velocityDecompSource$ can be written in terms of $\veloFlux$ and $\veloSource$ such that 
\begin{align*}
	\velocityDecompFlux(\state,t;\param) &= \conserveOp \veloFlux(\state,t;\param),\\
	\velocityDecompSource(\state,t;\param) &= \conserveOp \veloSource(\state,t;\param),
\end{align*}
and, thus, 
\begin{equation*}
	\velocityDecompSymbol(\states,t;\param) = \conserveOp \velo(\states,t;\param).
\end{equation*}
The subdomain conservation can be expressed as 
\begin{equation}\label{eq:govern_eq_C}
	\conserveOp\difftime{\states} = \conserveOp\velo(\states,t;\param).
\end{equation}
Applying a linear multistep scheme to \eqref{eq:govern_eq_C} yields 
\begin{equation}\label{eq:eq:fomODeltaEC}
\conserveOp\res^{n}(\state^{n};\param) = \zero.
\end{equation}
For theoretical aspects of this decomposition, we refer 
readers to Ref.~\cite{carlberg2018conservative}.

\begin{remark}[Lack of conservation  for Deep LSPG]\label{rem:lackCons}	
    We note that the Deep LSPG dynamics model \eqref{eq:disc_opt_problem} in general violates the conservation laws underlying the dynamical system of interest. This occurs because the objective function in \eqref{eq:disc_opt_problem} is generally nonzero at the solution, and thus conservation condition \eqref{eq:eq:fomODeltaEC} is not generally satisfied for \textit{any} decomposed  mesh $\meshDecomp$. This lack of conservation can lead to spurious generation or dissipation of physical quantities that should be conserved in principle (e.g., mass, momentum, energy).
\end{remark}

We aim to remedy this primary shortcoming of Deep LSPG with the proposed Deep Conservation projection method. In particular, we define the Deep Conservation dynamics model by equipping the nonlinear least-squares problem \eqref{eq:disc_opt_problem} with \textit{nonlinear equality constraints} corresponding to Eq.~\eqref{eq:eq:fomODeltaEC}, which has the effect of explicitly enforcing conservation over the decomposed mesh $\meshDecomp$. In particular, the Deep Conservation dynamics model computes latent states $\rdstate^n(\param)$, $n=1,\ldots,\ntime$ that satisfy 
\begin{equation}\label{eq:constrainedLSPG}
\begin{split}
	\underset{\grdstate\in\RR{\nstatered}}{\text{minimize}} &\left\|
	\res^{n}\left(
	\stateRef(\param)+\rdbasisnl(\grdstate);\param\right)\right\|_2^2\\
	\text{ subject to } &\conserveOp \res^{n}\left(\stateRef(\param)+\rdbasisnl(\grdstate);\param)\right) = \zero.
\end{split}
\end{equation}

To solve the problem \eqref{eq:constrainedLSPG} at each time instance, we follow the approach\footnote{In the original formulation, there is no  scalar for the update of Lagrange multipliers (i.e., $\LagrangeScalar=1$). In most of our numerical experiments, we follow this approach ($\LagrangeScalar=1$) unless otherwise specified.} 
considered in \cite{carlberg2018conservative} 
and apply sequential quadratic programming (SQP) with the Gauss--Newton Hessian
approximation, which leads to iterations
\begin{equation*}
\begin{split}
	\begin{bmatrix}
		\rdtestbasis^{n}(\rdstate^{n(k)};\param)\tran\rdtestbasis^{n}(\rdstate^{n(k)};\param) &\frac{1}{{\LagrangeScalar}} \rdtestbasis^{n}(\rdstate^{n(k)};\param)\tran \conserveOp\tran\\
		\conserveOp \rdtestbasis^{n}(\rdstate^{n(k)};\param) & \zero
	\end{bmatrix}
	\begin{bmatrix}
		\delta \rdstate^{n(k)}\\
		{\LagrangeScalar} \delta \lagrangeMul^{n(k)}
	\end{bmatrix} \\
	= - 
	\begin{bmatrix} 
		\rdtestbasis^{n}(\rdstate^{n(k)};\param)\tran \res^{n}(\stateRef(\param)+\rdbasisnl(\rdstate^{n(k)});\param) \\ 	\conserveOp \res^{n}(\stateRef(\param)+\rdbasisnl(\rdstate^{n(k)});\param)  ) 
	\end{bmatrix}
\end{split}
\end{equation*}
\begin{equation*}
	\begin{bmatrix}
	 	\rdstate^{n(k+1)}\\
		 \lagrangeMul^{n(k+1)}
	\end{bmatrix} 
	= 
	\begin{bmatrix}
		 \rdstate^{n(k)}\\
		 \lagrangeMul^{n(k)}
	\end{bmatrix}
	+ \learningrate^{n(k)} 
	\begin{bmatrix}
		\delta \rdstate^{n(k)}\\
		{\LagrangeScalar} \delta \lagrangeMul^{n(k)}
	\end{bmatrix},
\end{equation*}
where $\lagrangeMul^{n(k)}\in\RR{\nddstate}$ denotes Lagrange multipliers at time instance $n$ and iteration $k$ and $\learningrate^{n(k)}\in\RR{}$ is the step length that can be chosen, e.g., to satisfy the strong Wolfe conditions to ensure global convergence to a local solution of \eqref{eq:constrainedLSPG}.
\begin{remark}[Conservative LSPG projection]
	 Conservative LSPG projection \cite{carlberg2018conservative}
	employs the same formulation \eqref{eq:constrainedLSPG}, but with  reference state $\stateRef(\param)=\stateInit(\param)$ and linear parameterization function $\rdbasisnl:\drdstate\mapsto\rdbasis\drdstate$ as described in Remark \ref{rem:pod}.
\end{remark}

\section{Numerical experiments}\label{sec:numexp}
This section assesses the performance of (1) the proposed \DCNameCap\
projection, (2) \LSPGNameDeep\ projection, which also employs a nonlinear
embedding but does not enforce conservation, (3) POD--LSPG projection, which
employs a linear embedding and does not enforce conservation, and (4)
conservative LSPG projection, which employs a linear embedding but enforces
conservation.  We consider a parameterized Burgers' equation, as it is a
classical benchmark advection problem. We employ the numerical PDE tools and projection functionality provided by \texttt{pyMORTestbed} \cite{zahr2015progressive}, and we construct the autoencoder using \texttt{TensorFlow}  1.13.1 \cite{tensorflow2015-whitepaper}. 

The \DCNameCap\ and \LSPGNameDeep\ methods employ a 10-layer convolutional autoencoder. The encoder  $\encoder$ consists of $5$ layers  with $4$ convolutional layers, followed by $1$ fully-connected layer. The decoder $\decoder$ consists of 
$1$ fully-connected layer, followed by $4$ transposed-convolution layers.  The latent state  
is of dimension $\nstatered$, which will
vary during the experiments to define different latent-state dimensions.  The size of the convolutional kernels in the encoder and the decoder are $\{16,8,4,4\}$ and $\{4,4,8,16\}$; the numbers of kernel filters in each convolutional and transposed-convolutional layer are $\{8,16,32,64\}$ and $\{64,32,16,1\}$; the stride is configured as $\{2,4,4,4\}$ and $\{4,4,4,2\}$; the ``SAME'' padding
strategy is used; and no pooling is used. For the nonlinear activation functions, we use exponential linear units (ELU) \cite{clevert2015fast}, and no activation function in the output layer.

We consider a parameterized inviscid Burgers' equation~\cite{hirsch2007numerical}, where 
the system is governed by a conservation law of the form \eqref{eq:PDE}
with $\nConservation=1$, $\nDim=1$, $\spatialspace = [0,100]$, 
\begin{equation*}\label{eq:burg}
	\frac{d}{dt} \int_{\generalSubdomain}{\conservedQuantity{}}(\xoned,t;\param)d\xoned +\int_{\generalInterface}\frac{\eqvar(\xoned,t;\param)^2}{2}\cdot n(\xoned)ds = \int_{\generalSubdomain}0.02 e^{\paramelem{2} \xoned}d\xoned,
\end{equation*}
with initial and boundary conditions
$\eqvar(\xoned,0;\param) = 1, \forall \xoned \in [0,100]$,
$\eqvar(0,t;\param) = \paramelem{1},   \forall t \in [0,T]$.
There are $\nparam=2$ parameters (i.e., $\param \equiv(\paramelem{1},\paramelem{2})$) with the parameter domain $\paramspace=[4.25, 5.5] \times [0.015,0.03]$, and the final time is set to $T=35$.  We apply Godunov's scheme~\cite{hirsch2007numerical}, which is a finite-volume
scheme, with $\nControlVol=512$ control volumes, which results in a system of ODEs of the form \eqref{eq:govern_eq} with $\dofhf = 512$ spatial degrees of freedom. For time discretization, we use the backward-Euler scheme (i.e., $k=1$, $\alpha_0=\beta_0=1$, $\alpha_1=-1$, and $\beta_1=0$ in Eq.\
\eqref{eq:disc_res}). We consider a uniform time step $\timestep = 0.07$, resulting in $\nseq = 500$.

For offline training, we set the training-parameter instances to
$\paramspacetrain=\{(4.25 + (1.25/9)i,\ 0.015+(0.015/7)j)\}_{i=0,\ldots,9;\
j=0,\ldots 7}$, resulting in $\ntrain=80$ training-parameter instances. After collecting the snapshots, we split the snapshot matrix \eqref{eq:snapshotMat} into a training set and a validation set;  the fraction of snapshots to use for validation is 10\%. Then we compute optimal parameters $(\encparamOpt,\decparamOpt)$ using Adam optimizer \cite{kingma2014adam} with an initial uniform learning rate $\learningrate = 10^{-4}$ and initial parameters ($\encparam^{(0)},\decparam^{(0)}$) are computed via Xavier initialization \cite{glorot2010understanding}. The number of minibatches determined by a fixed batch size of 20; a maximum number of epochs is $\nepoch = 1550$; and early-stopping is enforced if
the loss on the validation set fails to decrease over 200 epochs. 

In the online-test stage, solutions of the model problem at a parameter instance $\paramtest^{1} =\left(4.3, 0.021\right) \notin \paramspacetrain$ are computed using all considered projection methods. The stopping criterion for all nonlinear solvers is the relative residual and the default stopping tolerance is $\tau=10^{-5}$. For conservative LSPG and \DCNameCap\ methods, we consider decomposed meshes, where the subdomains are defined such that they have
equal size ($|\subdomainArg{i}| = |\subdomainArg{j}|$, $\forall i,j$),
do not overlap ($\text{meas}(\subdomainArg{i}\cap\subdomainArg{j})=0$,
$\forall i\neq j$), and their union is equal to the full spatial
domain
($\cup_i\subdomainArg{i}=\spatialspace$).

\begin{figure}[!t]
    \centering
    \begin{minipage}{.45\linewidth}
        \centering
        \includegraphics[width=.45 \linewidth]{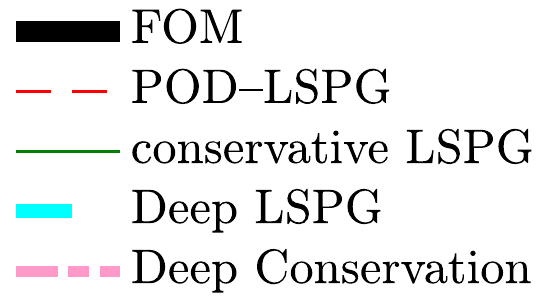}
    \end{minipage}
    \begin{minipage}{.45\linewidth}
        \subfloat[Full-order model solutions]{\label{fig:burger_fom} \includegraphics[width=1.0 \linewidth]{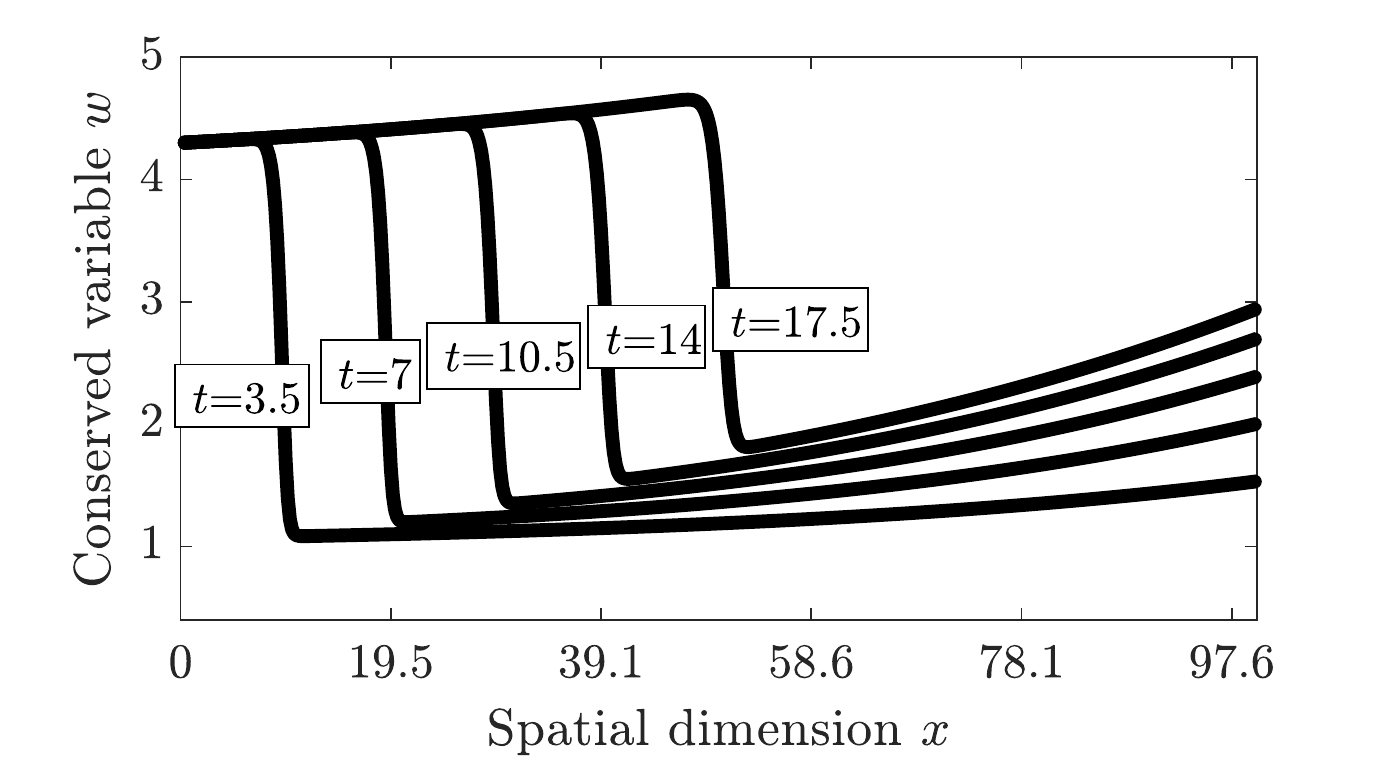}}
    \end{minipage}\\
    \begin{minipage}{.45\linewidth}
        \subfloat[Latent dimension,  $\dofrom\!=\!2$]  {\label{fig:burger_snapshot_p2}\includegraphics[width=1.0 \linewidth]{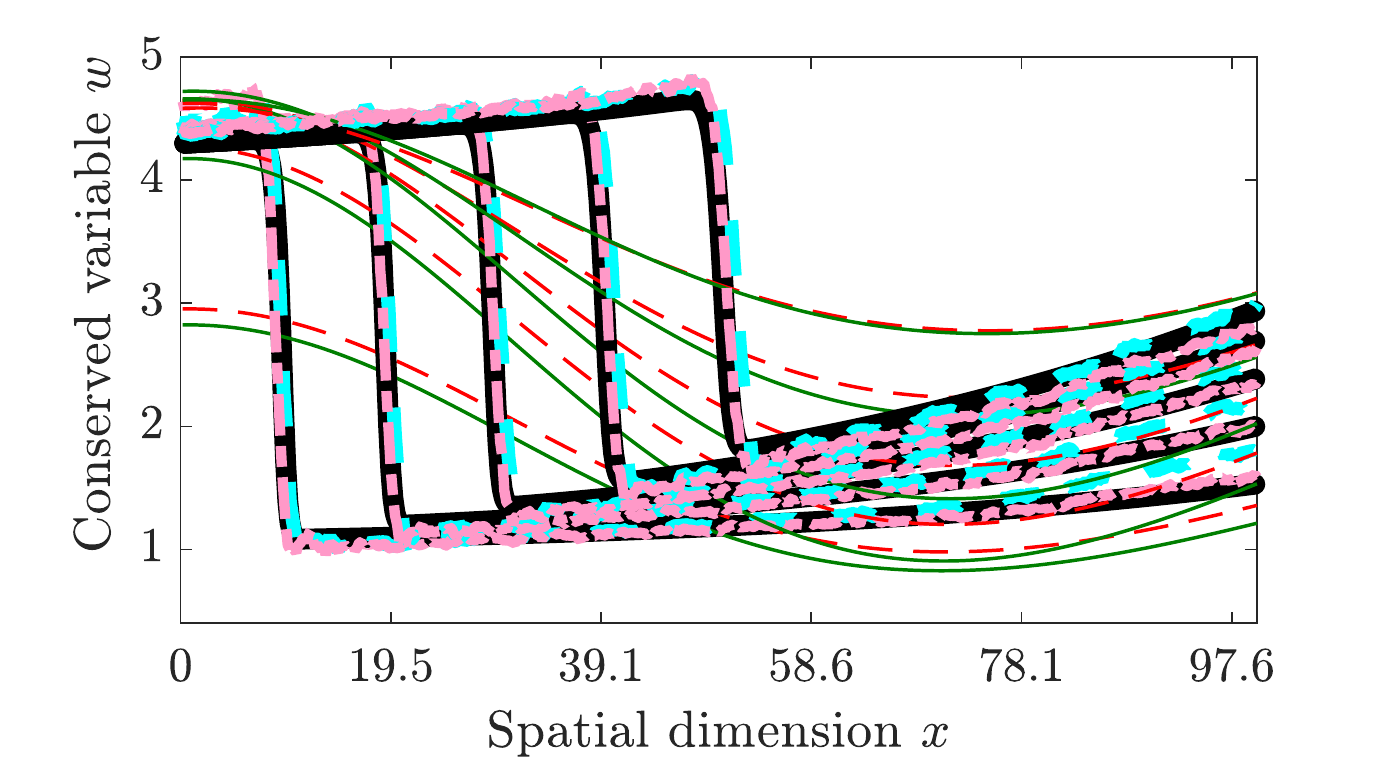}}
    \end{minipage}
    \begin{minipage}{.45\linewidth}
        \subfloat[Latent dimension,  $\dofrom\!=\!4$]  {\label{fig:burger_snapshot_p4}\includegraphics[width=1.0 \linewidth]{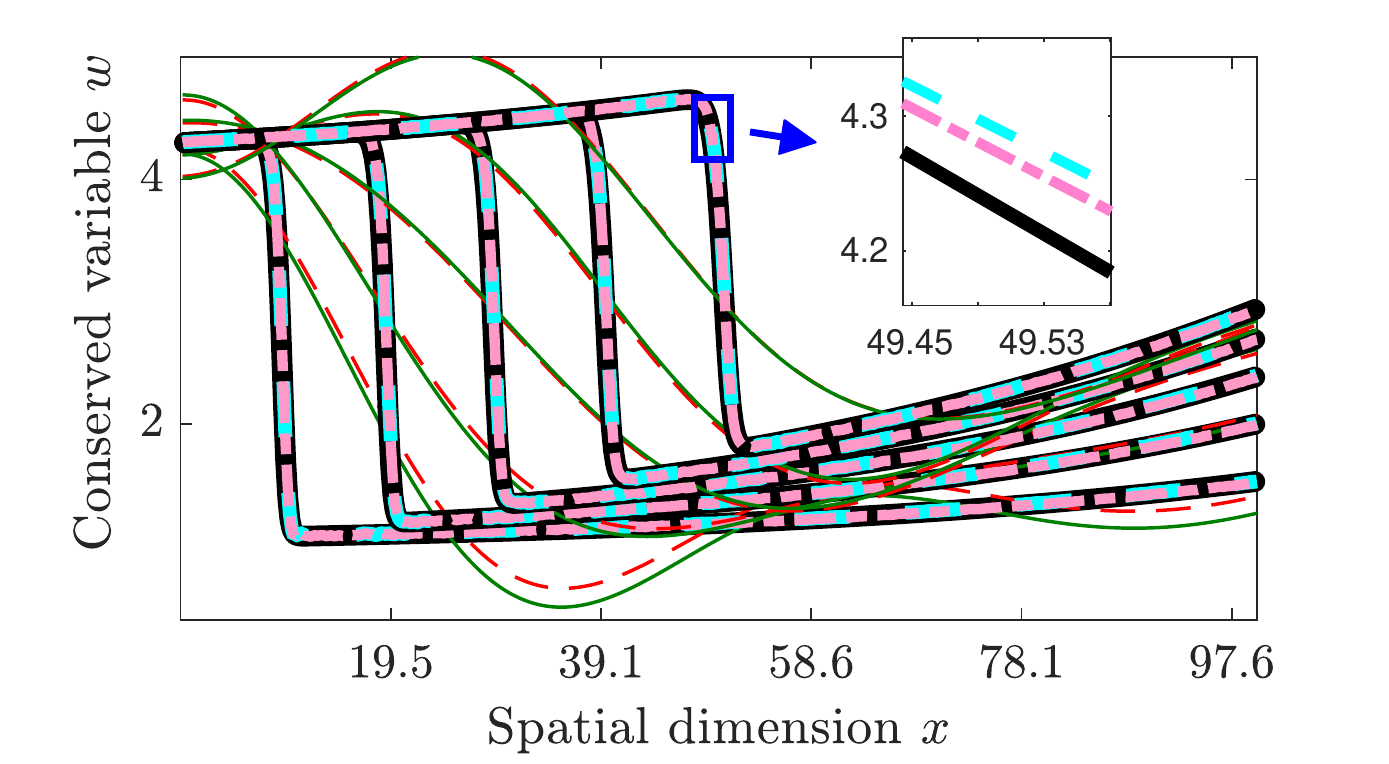}}
    \end{minipage}
    \caption{Online solutions at time
	instances $t=\{3.5, 7.0, 10.5, 14, 17.5\}$ computed by the FOM, POD--LSPG,
	conservative LSPG, \LSPGNameDeep, and \DCNameCap. All conservative
	methods enforce global conservation, i.e., they employ $\nSubdomains=1$ subdomain with $\subdomainArg{1}=\spatialspace$.}
	\label{fig:burger_snapshot_fig} 
\end{figure}

Figure \ref{fig:burger_snapshot_fig} reports solutions at five different time instances computed using FOM and all of the considered projection methods.  Figure \ref{fig:burger_fom} shows the FOM solutions demonstrating that the location of the shock, where the discontinuity exists, moves from left to right as time evolves. All projection methods employ the same latent-state dimension of  $\dofrom=2$ and $\dofrom=4$. These results clearly demonstrate that the projection-based methods using nonlinear embeddings yield significantly lower error than the methods using the classical linear embeddings.  Moreover, Figure \ref{fig:burger_snapshot_fig} demonstrates that the accuracy of the nonlinear embedding solutions is significantly improved as the latent dimension is increased from $\dofrom=2$
(Figure  \ref{fig:burger_snapshot_p2}) to $\dofrom=4$ (Figure \ref{fig:burger_snapshot_p4}).  As the solutions of the problem are characterized by three factors $(t,\paramelem{1},\paramelem{2})$, the intrinsic solution-manifold dimension is (at most) 3. Thus, autoencoders with the latent dimension larger than or equal to $\dofrom = 3$ will be able to reconstruct the original input data given sufficient capacity.

\begin{figure}[!t]
    \centering
    \begin{minipage}[t]{.31\linewidth}
        \centering
        \subfloat[State error $\stateError$, $\hybridTerm=0$]  {\includegraphics[width=.95 \linewidth]{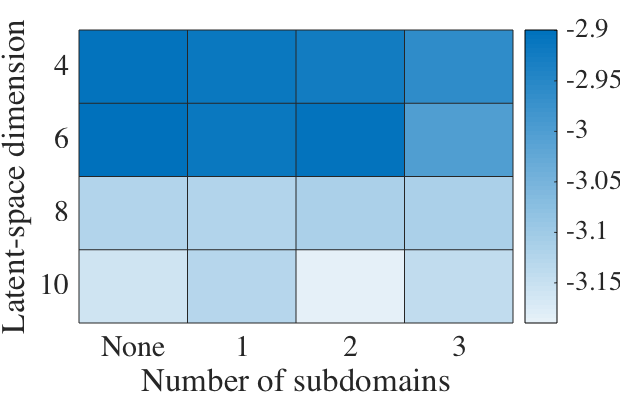}
        \label{fig:state_error_deep}}
    \end{minipage}
    \begin{minipage}[t]{.31\linewidth}    
        \centering
	    \subfloat[Global conserved-variable error $\stateErrorGlobal$, $\hybridTerm\!=\!0$]  {\includegraphics[width=.95 \linewidth]{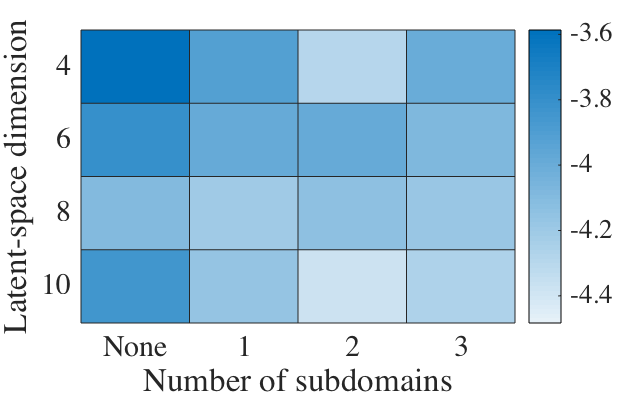} \label{fig:state_globalerror_deep}}
    \end{minipage}
    \begin{minipage}[t]{.31\linewidth}
        \centering
	    \subfloat[Global conservation violation $\resErrorGlobal$,  $\hybridTerm=0$]  {\includegraphics[width=.95 \linewidth]{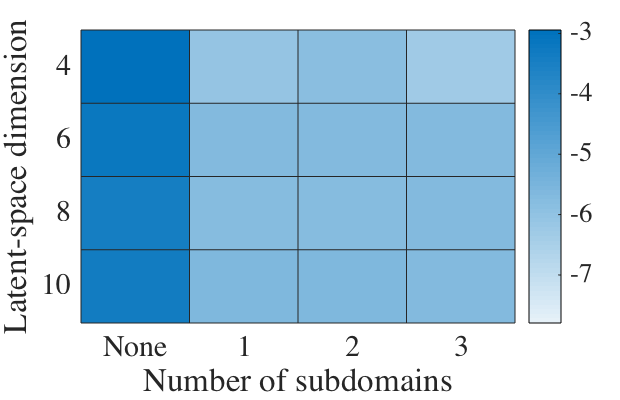} \label{fig:state_globalconserve_deep}}\\
    \end{minipage}\\
    \begin{minipage}[t]{.31\linewidth}
        \centering
        \subfloat[State error $\stateError$, $\hybridTerm=1$]  {\includegraphics[width=.95 \linewidth]{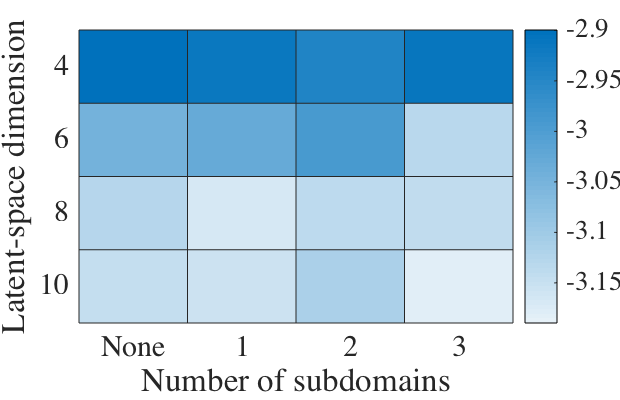}
        \label{fig:state_error_dc}}
    \end{minipage}
    \begin{minipage}[t]{.31\linewidth}    
        \centering
	    \subfloat[Global conserved-variable error $\stateErrorGlobal$, $\hybridTerm\!=\!1$]  {\includegraphics[width=.95 \linewidth]{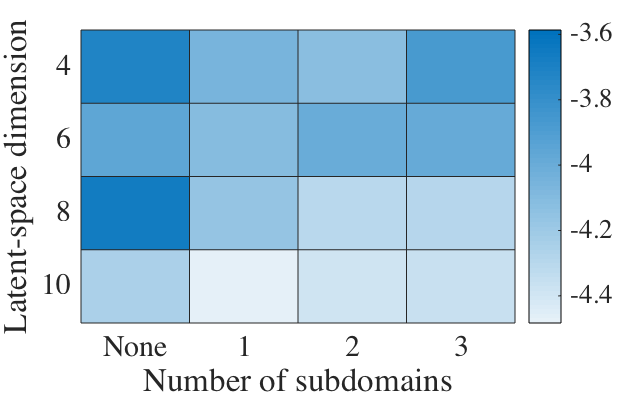} \label{fig:state_globalerror_dc}}
    \end{minipage}
    \begin{minipage}[t]{.31\linewidth}
        \centering
	    \subfloat[Global conservation violation $\resErrorGlobal$, $\hybridTerm=1$]  {\includegraphics[width=.95 \linewidth]{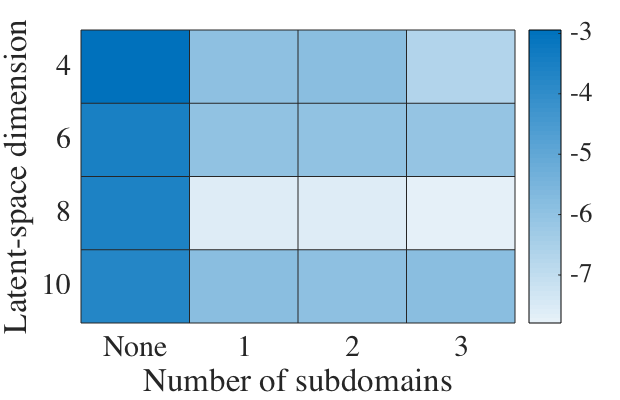} \label{fig:state_globalconserve_dc}}
    \end{minipage}
    \subfloat[Proportion of error metrics where \DCNameCap\ with $\rho=1$ outperforms \DCNameCap\ with $\rho=0$ in 1, 2, and all 3 metrics.]  {\includegraphics[width=.5 \linewidth]{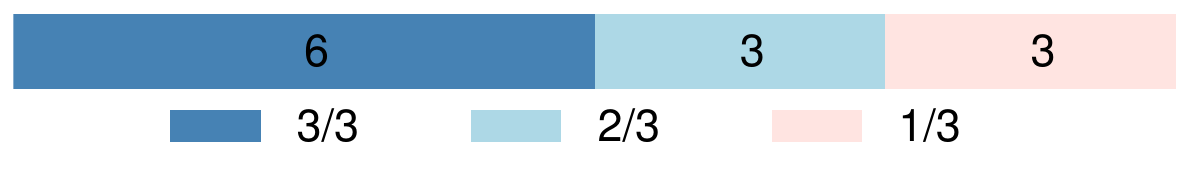}
    \label{fig:barchart}}
    \caption{Error metrics in $\log_{10}$ scale for \LSPGNameDeep\ (None) and \DCNameCap\ ($\nSubdomains \geq 1$) for varying latent-space dimensions $\nstatered$ (vertical axis) and for varying numbers of subdomains $\nSubdomains$ (horizontal axis) with the baseline ($\hybridTerm\!=\!0$, left) and the hybrid ($\hybridTerm\!=\!1$, right) training objective functions.}
	\label{fig:error}
\end{figure}

Now, we quantitatively assess the accuracy of of the approximated state
$\aprxstate$ computed using \LSPGNameDeep\ and \DCNameCap\ methods with the following
metrics: 
1) the state error,
 $$\stateError\!\! \defeq \!\! \sqrt{ \left. \sum_{n=1}^{\nseq}  \|\states^n(\param)-\aprxstate^n(\param) \|_2^2 \middle/ \sum_{n=1}^{\nseq}\|\states^n(\param)\|_2^2\right.} ,$$ 
2)~the error in the globally conserved variables, 
$$\stateErrorGlobal\!~\defeq\!~ \sqrt{\left. \!\sum_{n=1}^{\nseq} \!\! \| \conserveOpGlobal (\states^n(\param)\!-\!\aprxstate^n(\param)) \|_2^2\! \middle/ \! \sum_{n=1}^{\nseq} \!\!	\|\conserveOpGlobal\states^n(\param)\|_2^2\right.} ,$$ 
and 3) the violation in global conservation, 
$$\resErrorGlobal \defeq  \sqrt{\sum_{n=1}^{\nseq}  \| \conserveOpGlobal \res^{n}(\aprxstate^{n}(\param);\param) \|_2^2},$$ where $\conserveOpGlobal\in\RRplus^{\nConservation\times\ndof}$ is the
operator $\conserveOp$ associated with the global conservation
$\meshDecompGlobal \defeq
\{\spatialspace\}$. 

Figure \ref{fig:error} reports these quantities for the
considered methods. 
\textbf{These results illustrate that the best performance in most cases is
obtained through the combination of a nonlinear embedding and conservation
enforcement as provided by the proposed Deep Conservation method}. That is, lower errors can be achieved by using the proposed \DCNameCap\ than the \LSPGNameDeep\ projection. In particular, \DCNameCap\ reduces the global conservation violation $\resErrorGlobal$ by more than an order of magnitude relative to that of \LSPGNameDeep. As numerically demonstrated in \cite{carlberg2018conservative}, minimizing the residual with the conservation constraint leads to smaller errors in states and globally conserved states (Figures \ref{fig:state_error_deep}--\ref{fig:state_globalerror_deep} and \ref{fig:state_error_dc}--\ref{fig:state_globalerror_dc}).

\begin{figure}[!t]
    \centering
    \begin{minipage}{.45\linewidth}
        \subfloat[Global conservation violation
    $\resErrorGlobal^n$, $\hybridTerm=0$]  {\includegraphics[width=.9 \linewidth]{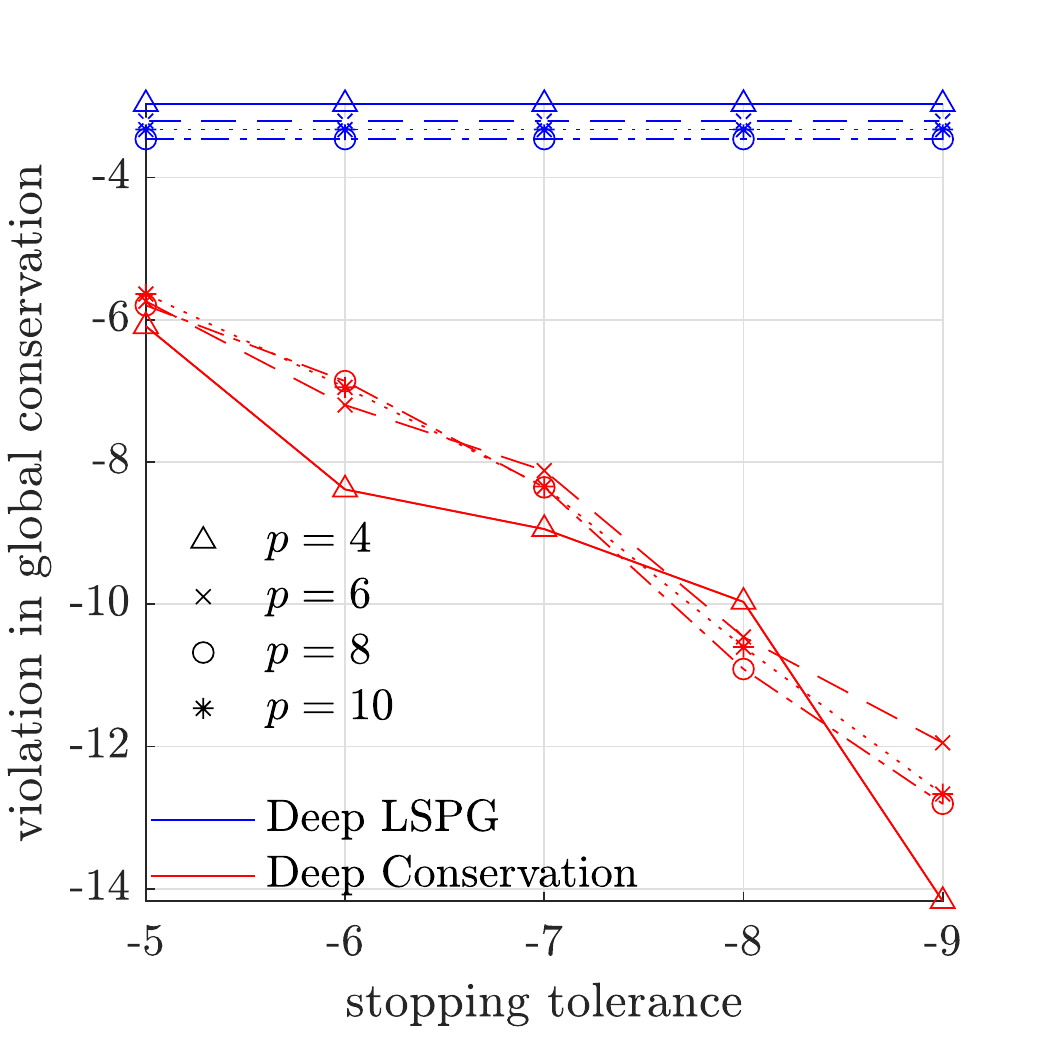}}
    \end{minipage}
    \begin{minipage}{.45\linewidth}
			\subfloat[Global conservation violation
    $\resErrorGlobal^n$, $\hybridTerm=1$]  {\includegraphics[width=.9 \linewidth]{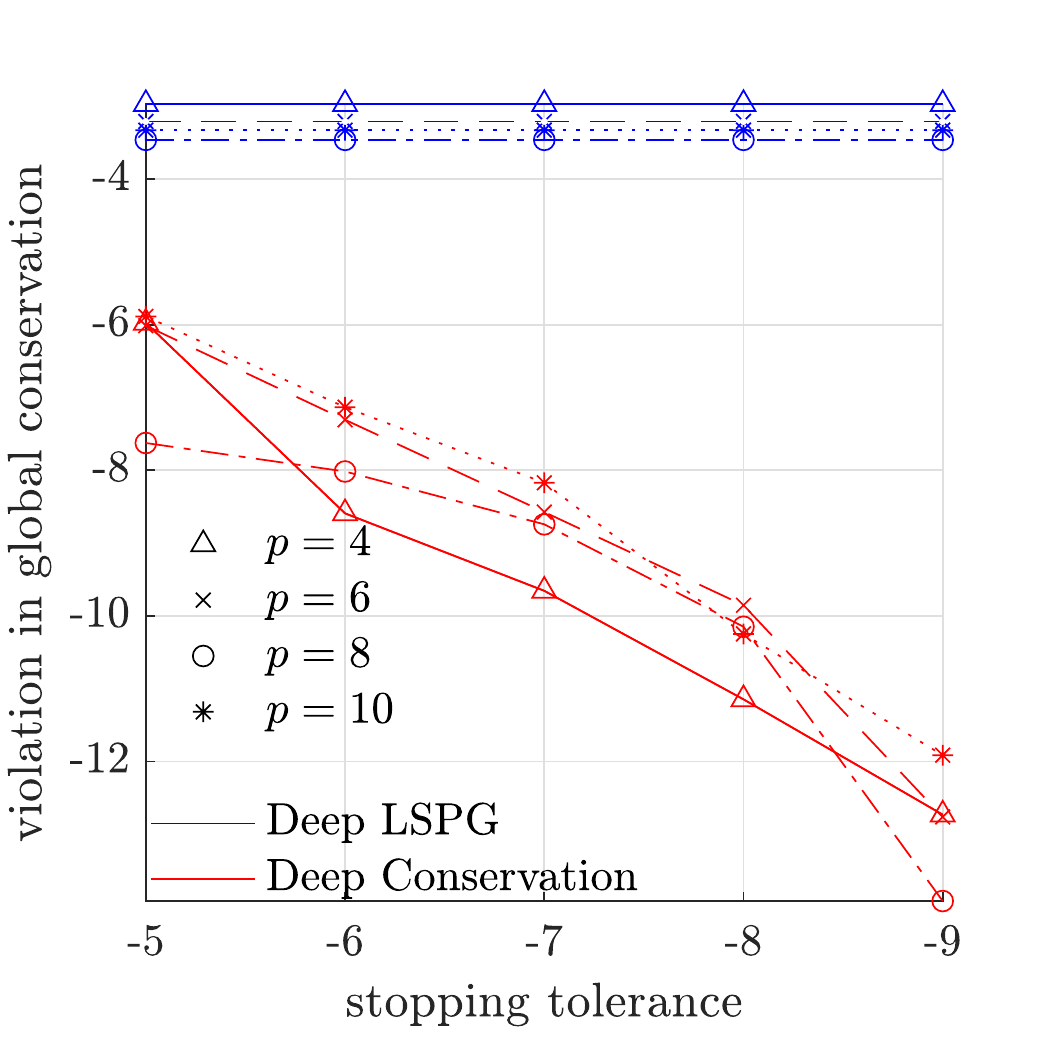}}
    \end{minipage}
    \caption{Global conseration violation 
    $\resErrorGlobal^n$ for \LSPGNameDeep\ and \DCNameCap\ ($\nSubdomains = 1$) for varying stopping tolerance of the nonlinear solvers.}
	\label{fig:violation_vs_tol}
\end{figure}

Figure \ref{fig:error} also shows that \DCNameCap\ with the hybrid autoencoder objective function ($\penaltyTerm=1$, right) can lead to smaller errors than \DCNameCap\ with the baseline autoencoder objective function ($\penaltyTerm=0$, left) . The hybrid objective function helps improving the accuracy in terms of violation in global conservation (Figures \ref{fig:state_globalconserve_deep}--\ref{fig:state_globalconserve_dc}). 
Based on the 12 experimental settings used in Figure \ref{fig:error} (i.e., combinations of $p =\{4,6,8,10\}$ and $\nSubdomains=\{1,2,3\}$), Figure \ref{fig:barchart} reports the proportions of the error metrics where the \DCNameCap\ with the hybrid objective ($\penaltyTerm=1$) outperforms \DCNameCap\ with the baseline objective ($\penaltyTerm=0$) in 1, 2, and, all 3 error metrics.

Figure \ref{fig:violation_vs_tol} further highlights the substantial
performanc improvements offered by the proposed  \DCNameCap\ method
(associated with solving \eqref{eq:constrainedLSPG} online) over
\LSPGNameDeep\ (associated with solving \eqref{eq:disc_opt_problem} online).
In particular, this figure illustrates that
having a stringent stopping tolerance for \DCNameCap\ leads to decrease in the
global conservation violation $\resErrorGlobal$, whereas 
\LSPGNameDeep\  fails to improve $\resErrorGlobal$ even with stringent
stopping tolerances.\footnote{Note that the state error, $\stateError$, and the error in the globally conserved variables, $\stateErrorGlobal$,
obtained for varying tolerances are not reported as there are only negligible
changes.}

\begin{figure}[!t]
    \centering
    \begin{minipage}{.19\linewidth}
    \includegraphics[width=.9 \linewidth]{figs/lgd.pdf}
    \end{minipage}
    \begin{minipage}{.8\linewidth}
    \subfloat[Test parameter instance (5.15, 0.0285)]  {\includegraphics[width=.5 \linewidth]{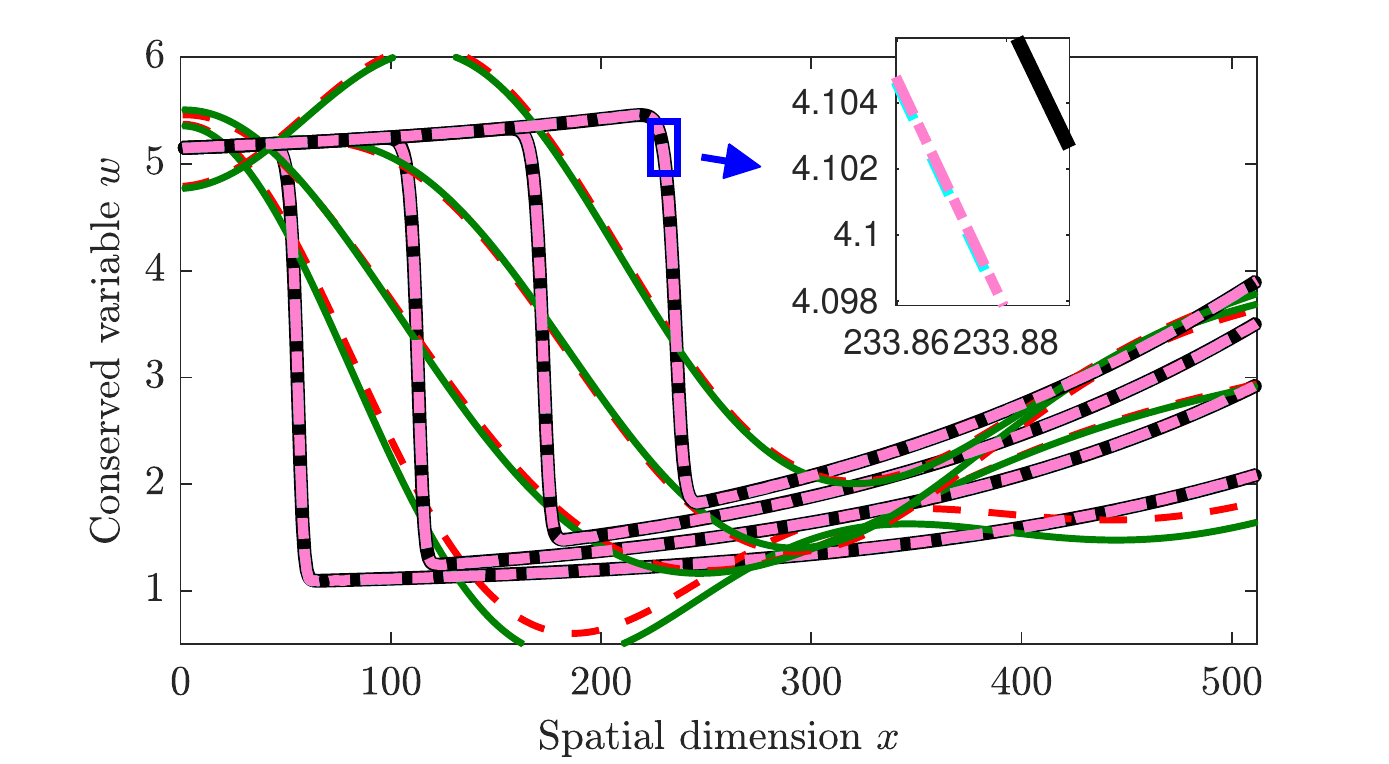}}
    \subfloat[Test parameter instance (4.1, 0.0315)]  {\includegraphics[width=.5 \linewidth]{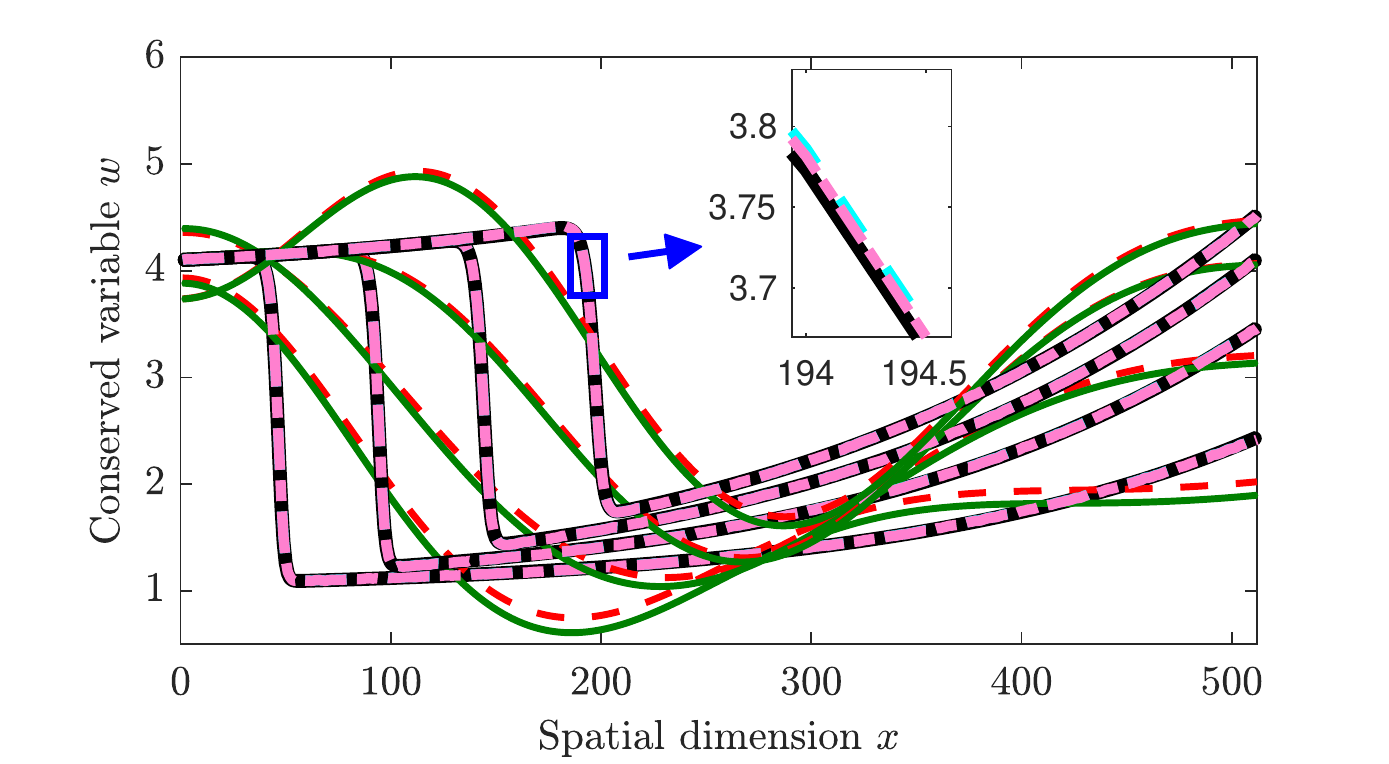}}
    \end{minipage}
    \caption{Online solutions at time
	instances $t=\{3.5, 7.0, 10.5, 14\}$ computed by the FOM, POD--LSPG,
	conservative LSPG, \LSPGNameDeep\, and \DCNameCap. All conservative
	methods employ $\nSubdomains=1$ subdomains with the autoencoder of the latent dimension $\dofrom=4$.}
	\label{fig:burger_snapshot_additional_fig} 
\end{figure}

We continue the same numerical experiments on two more test parameter instances, $\paramtest^2=(5.15, 0.0285)$ and $\paramtest^3=(4.1, 0.0315)$, where both parameter instances are not included in the training dataset $\paramspacetrain$. Note that the third test parameter instance $\paramtest^3=(4.1, 0.0315)$ is outside of the parameter domain (i.e., $(4.1, 0.0315) \notin \paramspace=[4.25, 5.5] \times [0.015,0.03]$). Figure \ref{fig:burger_snapshot_additional_fig} depicts the solutions at time instances $t=\{3.5, 7.0, 10.5, 14\}$ computed by using all considered methods: FOM, POD--LSPG, conservative LSPG, \LSPGNameDeep, and \DCNameCap. Figure \ref{fig:burger_snapshot_additional_fig} shows the solutions of the inviscid Burgers' equation for the two test-parameter instances ($\paramtest^2=(5.15, 0.0285)$ and $\paramtest^3=(4.1, 0.0315)$) at time instances $t=\{3.5, 7.0, 10.5, 14\}$ computed by using FOM and all considered projection methods. As observed in the experiments with the first test parameter instance, with the latent dimension is $\dofrom=4$, both \LSPGNameDeep\ and \DCNameCap\ produce very accurate approximate solutions while the classical linear subspace methods, POD--LSPG and 
conservative LSPG, produce inaccurate approximate solutions. The magnifying boxes show that \DCNameCap\ produce more accurate solutions   than \LSPGNameDeep.

\begin{figure}[!t]
    \centering
        \begin{minipage}[t]{.31\linewidth}
        \subfloat[State error $\stateError$, $\hybridTerm=0$]  {\includegraphics[width=.9 \linewidth]{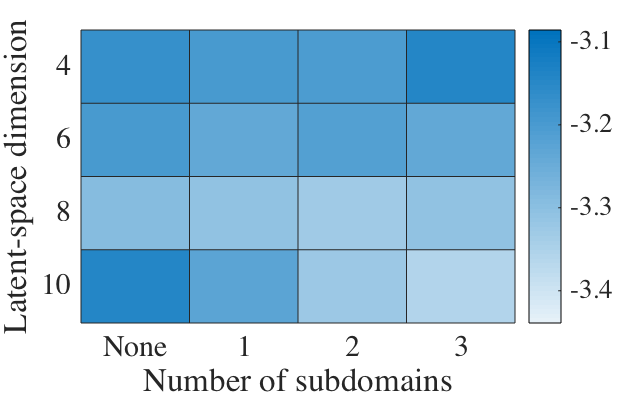}}
        \end{minipage}
        \begin{minipage}[t]{.31\linewidth}
	    \subfloat[Global conserved-variable error $\stateErrorGlobal$, $\hybridTerm=0$]  {\includegraphics[width=.9 \linewidth]{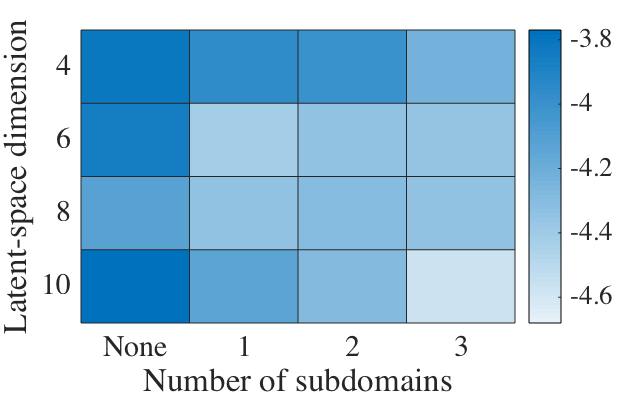}}
	    \end{minipage}
	    \begin{minipage}[t]{.31\linewidth}
	    \subfloat[Global conservation violation $\resErrorGlobal$, $\hybridTerm=0$]  {\includegraphics[width=.9 \linewidth]{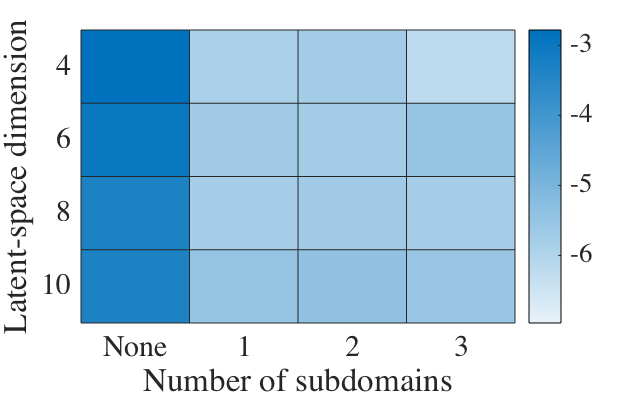}}
	    \end{minipage}
	    \\
	    \begin{minipage}[t]{.31\linewidth}
        \subfloat[State error $\stateError$, $\hybridTerm=1$]  {\includegraphics[width=.9 \linewidth]{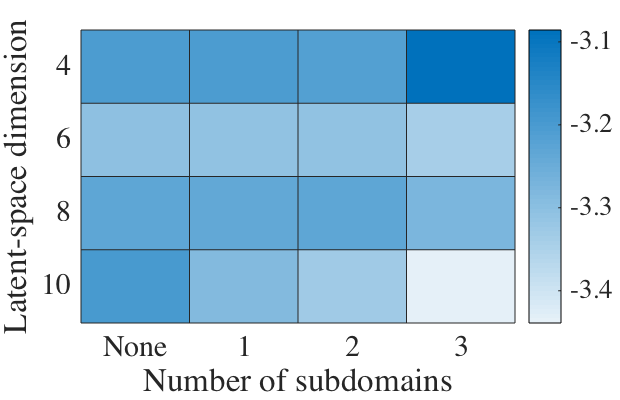}
        }
        \end{minipage}
        \begin{minipage}[t]{.31\linewidth}
	    \subfloat[Global conserved-variable error $\stateErrorGlobal$, $\hybridTerm=1$]  {\includegraphics[width=.9 \linewidth]{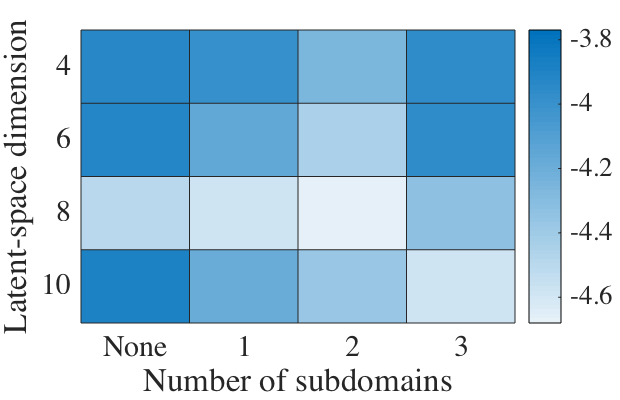}
	    }
	    \end{minipage}
	    \begin{minipage}[t]{.31\linewidth}
	    \subfloat[Global conservation violation $\resErrorGlobal$, $\hybridTerm=1$]  {\includegraphics[width=.9 \linewidth]{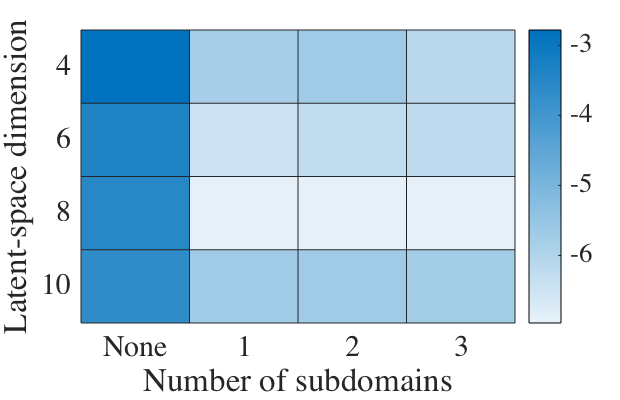}}
	    \end{minipage}
    
    \caption{Test parameter instance $\paramtest^2=(5.15, 0.0285)$: Error metrics in $\log_{10}$ scale for \LSPGNameDeep\ (None) and \DCNameCap\ ($\nSubdomains \geq 1$) for varying latent-space dimensions $\nstatered$ (vertical axis) and for varying numbers of subdomains $\nSubdomains$ (horizontal axis) with the baseline ($\hybridTerm=0$, top) and the hybrid ($\hybridTerm=1$, bottom) objective function.}
	\label{fig:error2}
\end{figure}

\begin{figure}[!t]
    \centering
        \begin{minipage}[t]{.31\linewidth}
        \subfloat[State error $\stateError$, $\hybridTerm=0$]  {\includegraphics[width=.9 \linewidth]{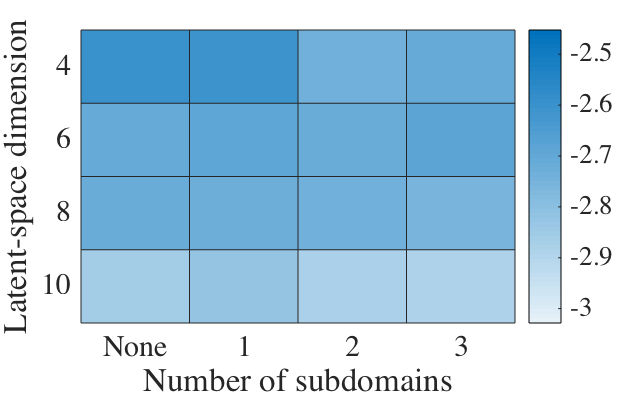}}
        \end{minipage}
        \begin{minipage}[t]{.31\linewidth}
	    \subfloat[Global conserved-variable error $\stateErrorGlobal$, $\hybridTerm=0$]  {\includegraphics[width=.9 \linewidth]{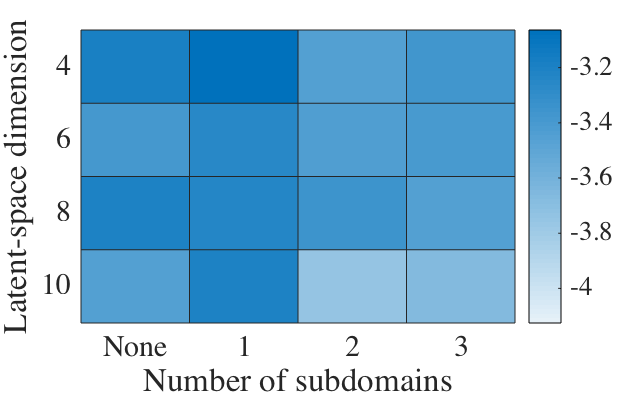}
	    } 
	    \end{minipage}
	    \begin{minipage}[t]{.31\linewidth}
	    \subfloat[Global conservation violation $\resErrorGlobal$, $\hybridTerm=0$]  {\includegraphics[width=.9 \linewidth]{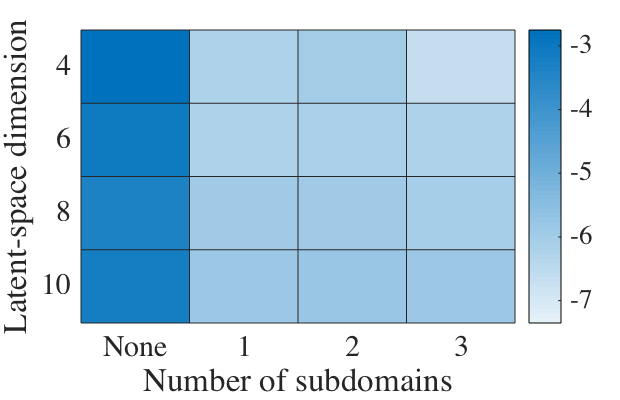}}
	    \end{minipage}
	    \\
	    \begin{minipage}[t]{.31\linewidth}
        \subfloat[State error $\stateError$, $\hybridTerm=1$]  {\includegraphics[width=.9 \linewidth]{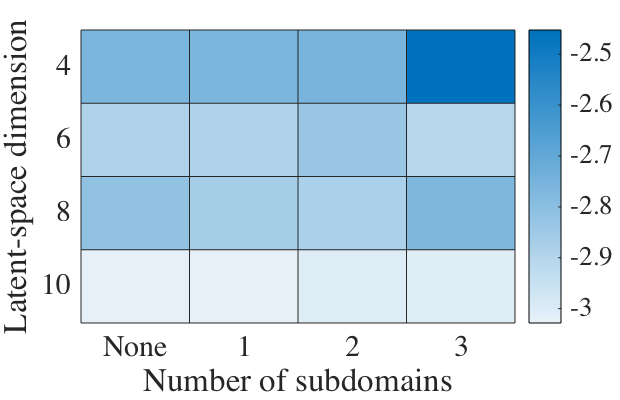}
        } 
        \end{minipage}
        \begin{minipage}[t]{.31\linewidth}
	    \subfloat[Global conserved-variable error $\stateErrorGlobal$, $\hybridTerm=1$]  {\includegraphics[width=.9 \linewidth]{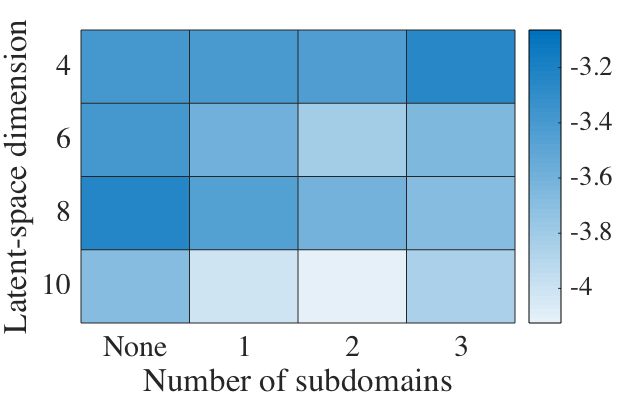}
	    } 
	    \end{minipage}
	    \begin{minipage}[t]{.31\linewidth}
	    \subfloat[Global conservation violation $\resErrorGlobal$, $\hybridTerm=1$]  {\includegraphics[width=.9 \linewidth]{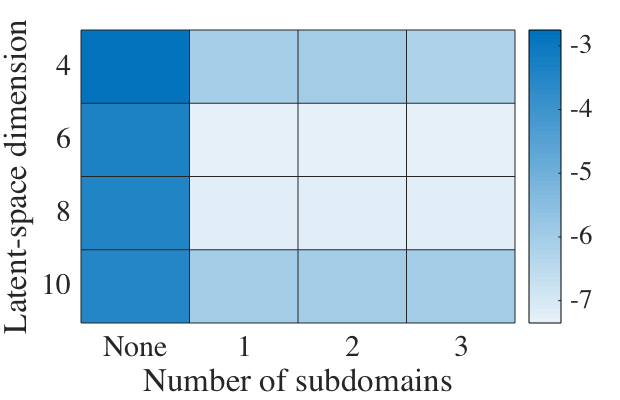}}
	    \end{minipage}
    \caption{Test parameter instance $\paramtest^3=(4.1, 0.0315)$: Error metrics in $\log_{10}$ scale for \LSPGNameDeep\ (None) and \DCNameCap\ ($\nSubdomains \geq 1$) for varying latent-space dimensions $\nstatered$ (vertical axis) and for varying numbers of subdomains $\nSubdomains$ (horizontal axis) with the baseline ($\hybridTerm=0$, top) and the hybrid ($\hybridTerm=1$, bottom) objective function.}
	\label{fig:error3}
\end{figure}

Figures \ref{fig:error2} and \ref{fig:error3} report three error metrics, the state error, the error in the globally conserved variables, and the violation in global conservation, for the test parameter instances $\paramtest^2=(5.15, 0.0285)$ and $\paramtest^3=(4.1, 0.0315)$, and again the best performance is obtained via \DCNameCap. In both test parameter instances, the violation in global conservation is improved by 2$\sim$3 orders of magnitude by using \DCNameCap\ compared to the results of \LSPGNameDeep. As observed in Section \ref{sec:numexp}, enforcing the conservation constraint again results in improvements in state errors and the error in the globally conserved variables in most cases. In particular, the error in the globally conserved variables are improved by an order of magnitude in many cases.

\begin{figure}[!t]
    \centering
    \begin{minipage}[t]{.45\linewidth}
    \subfloat[Test parameter instance $\paramtest^2=(5.15, 0.0285)$]  {\includegraphics[width=.9 \linewidth]{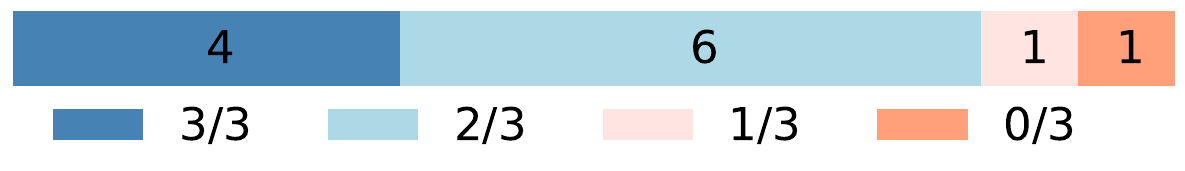}
    }
    \end{minipage}
    \begin{minipage}[t]{.45\linewidth}
    \subfloat[Test parameter instance $\paramtest^3=(4.1, 0.0315)$]  {\includegraphics[width=.9 \linewidth]{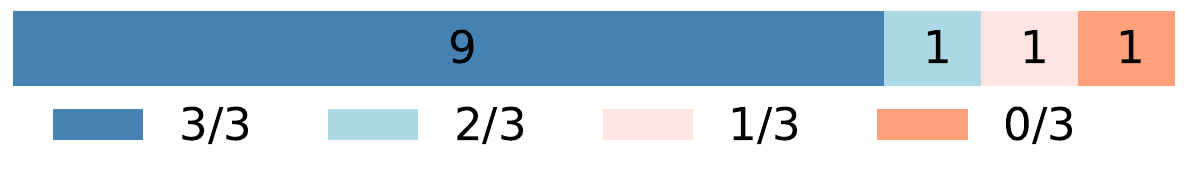}
    }
    \end{minipage}
    \caption{Proportion of error metrics where \DCNameCap\ with the hybrid objective function ($\hybridTerm=1$) outperforms \DCNameCap\ with the baseline objective function ($\hybridTerm=0$) in 1, 2, and all 3 metrics.}
    \label{fig:error_proportion}
\end{figure}

Figure \ref{fig:error_proportion} compares  \DCNameCap\ with the baseline autoencoder objective function ($\hybridTerm=0$) and with the the hybrid objective function ($\hybridTerm=1$). Again, as in the experiment with the first parameter instance, based on the 12 experimental settings used in Figures \ref{fig:error2}--\ref{fig:error3} (i.e., $p =\{4,6,8,10\}$ and $\nSubdomains=\{1,2,3\}$), Figure \ref{fig:error_proportion} reports the proportions of the error metrics where the \DCNameCap\ with $\hybridTerm=1$ outperforms \DCNameCap\ with $\hybridTerm=0$ in 1, 2, and, all 3 error metrics. We observe that adding the additional residual-minimizing objective function is helpful to improve the quality of the approximation as \DCNameCap\ with $\rho=1$ improves at least two error metrics in 10 experimental settings for both test parameter instances.

\begin{figure}[!h]
    \centering
    \subfloat[$p=4$]  {\includegraphics[width=.5 \linewidth]{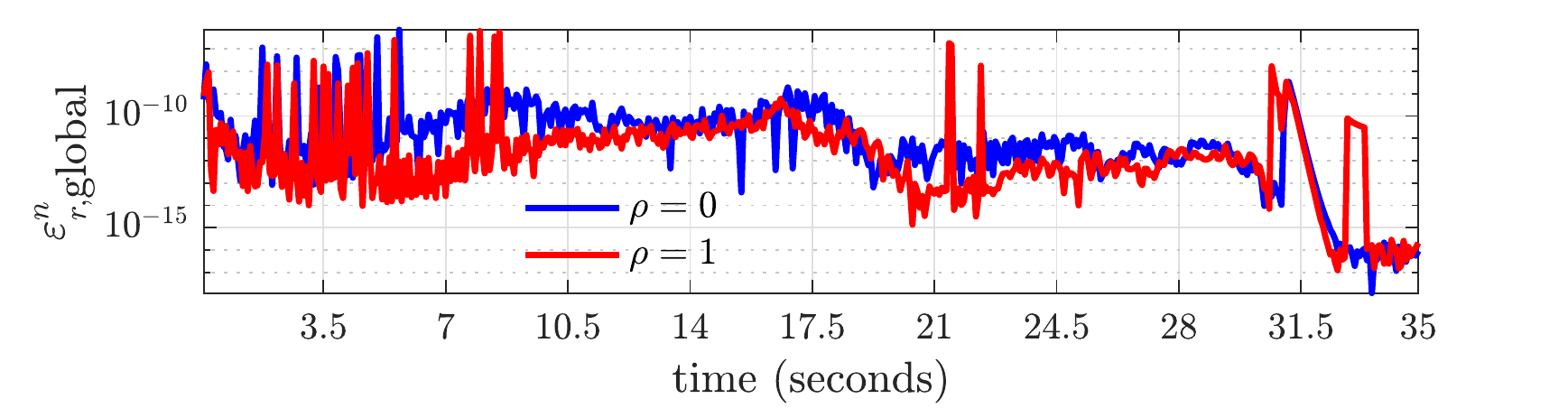}}
    \subfloat[$p=6$]  {\includegraphics[width=.5 \linewidth]{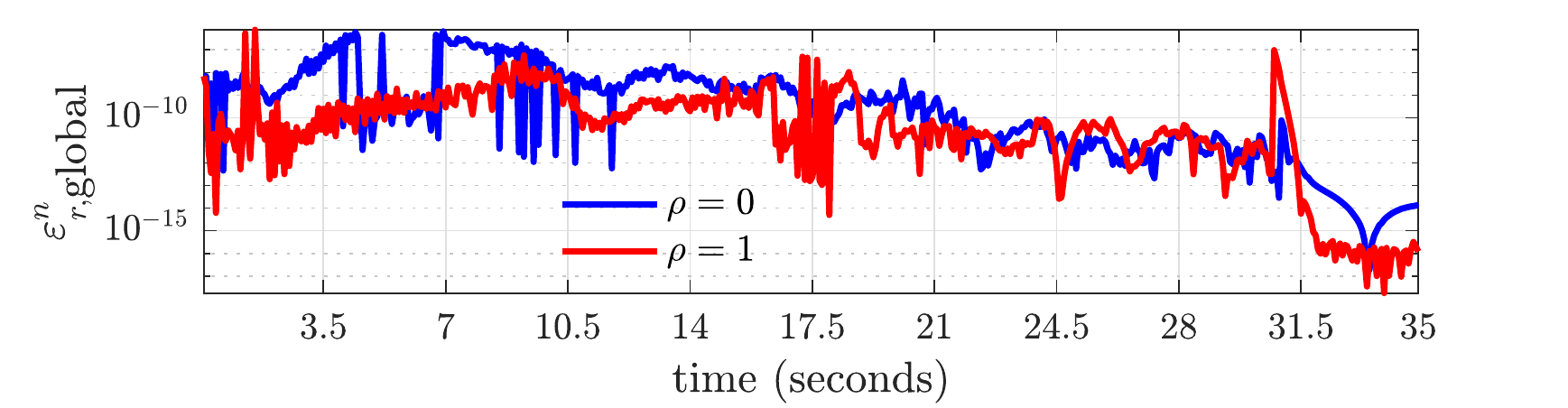}}\\
    \subfloat[$p=8$]  {\includegraphics[width=.5 \linewidth]{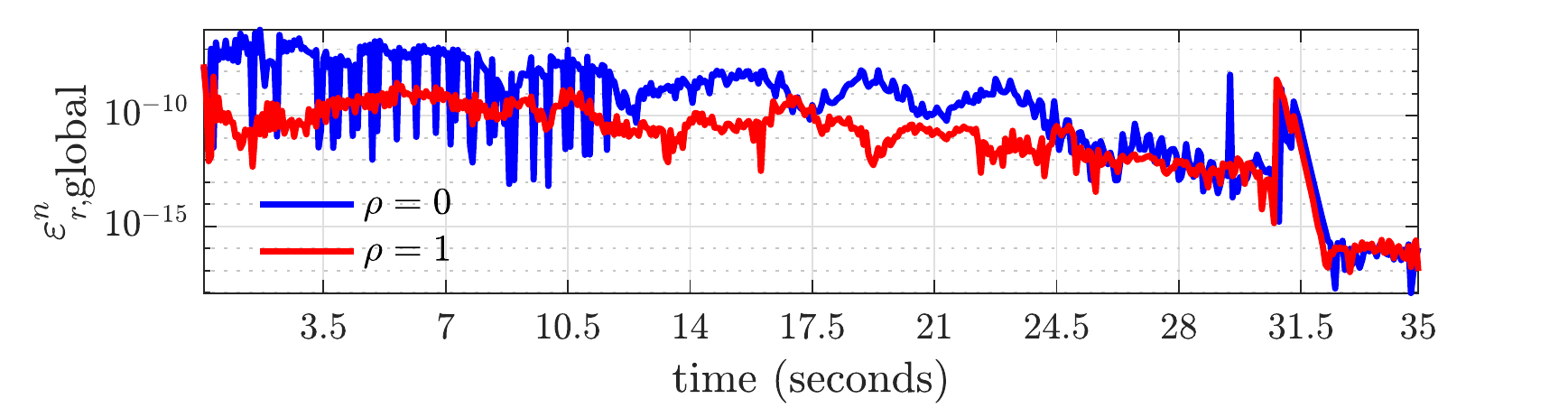}}
    \subfloat[$p=10$]  {\includegraphics[width=.5 \linewidth]{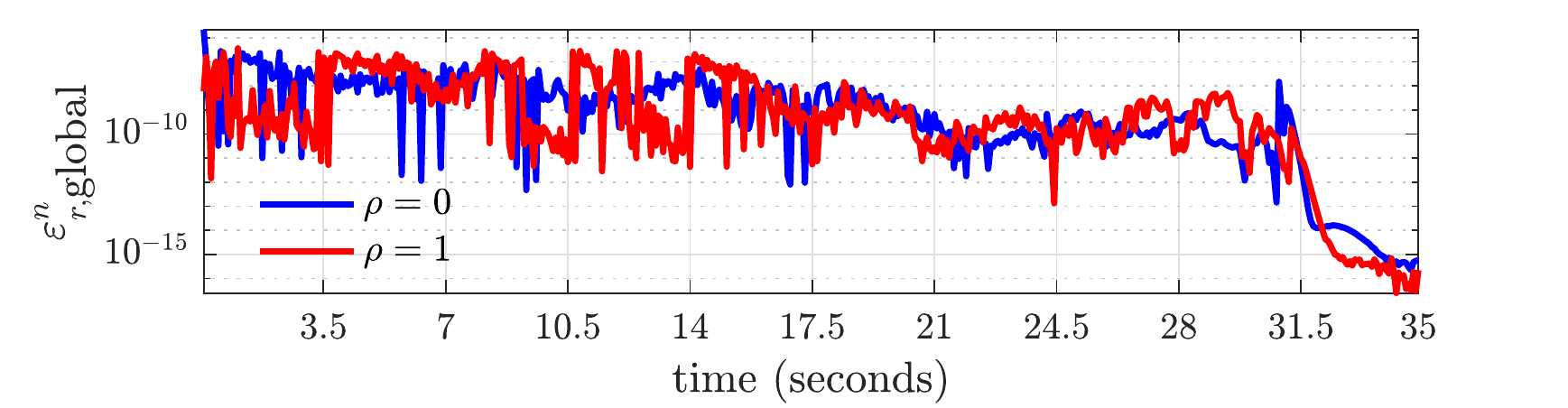}}
    \caption{The time evolution of the time-instantaneous violation in global conservation
    $\resErrorNGlobal^n$ computed using \DCNameCap\ ($\nSubdomains=1$) for varying latent-space dimensions $\nstatered$
	 with the autoencoders objective function with $\hybridTerm=0$ (blue) and  $\hybridTerm=1$ (red).}
	\label{fig:violation_instances}
\end{figure}

The performances of \DCNameCap\ with $\penaltyTerm=0$ and $\penaltyTerm=1$ are further assessed with an additional error metric,  time-instantaneous violation in global conservation
$\resErrorNGlobal^n \defeq \| \conserveOpGlobal \res^{n}(\aprxstate^{n}(\param);\param) \|_2^2$, $n=1,\ldots,\nseq$.
Figure \ref{fig:violation_instances} reports $\resErrorNGlobal^n$ and the results illustrate that \DCNameCap\ with $\penaltyTerm=1$ produces more accurate  approximate states in terms of violation of global conservation $\resErrorNGlobal^n$. 

\begin{figure}[!h]
    \centering
    \subfloat[$p=4$, $\LagrangeScalar=1$]  {\includegraphics[width=.5 \linewidth]{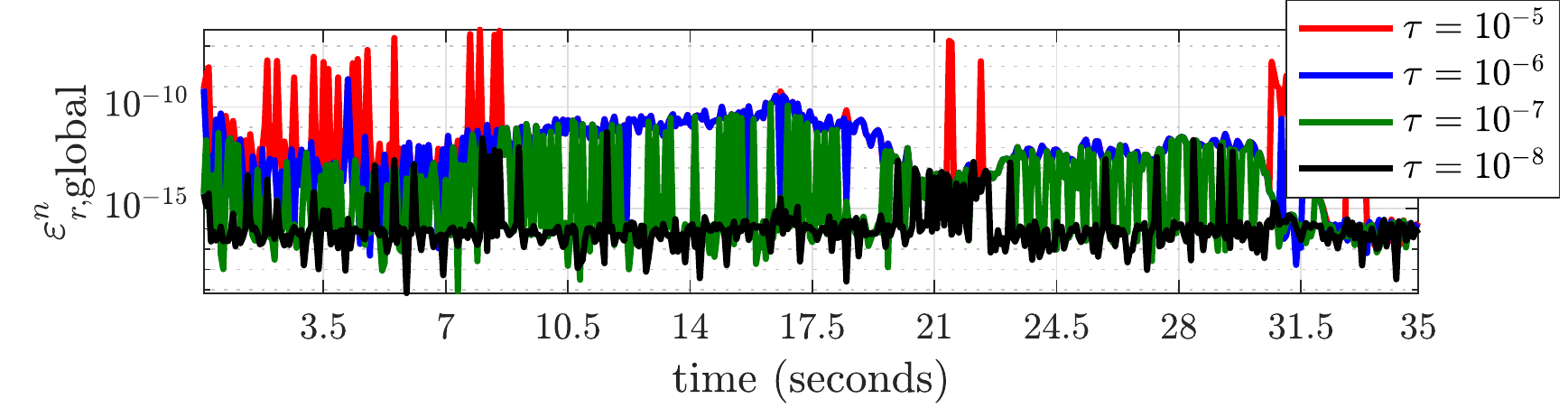}\label{fig:res_varying_tol}}
    \subfloat[$p=4$, $\tau=10^{-5}$]  {\includegraphics[width=.5 \linewidth]{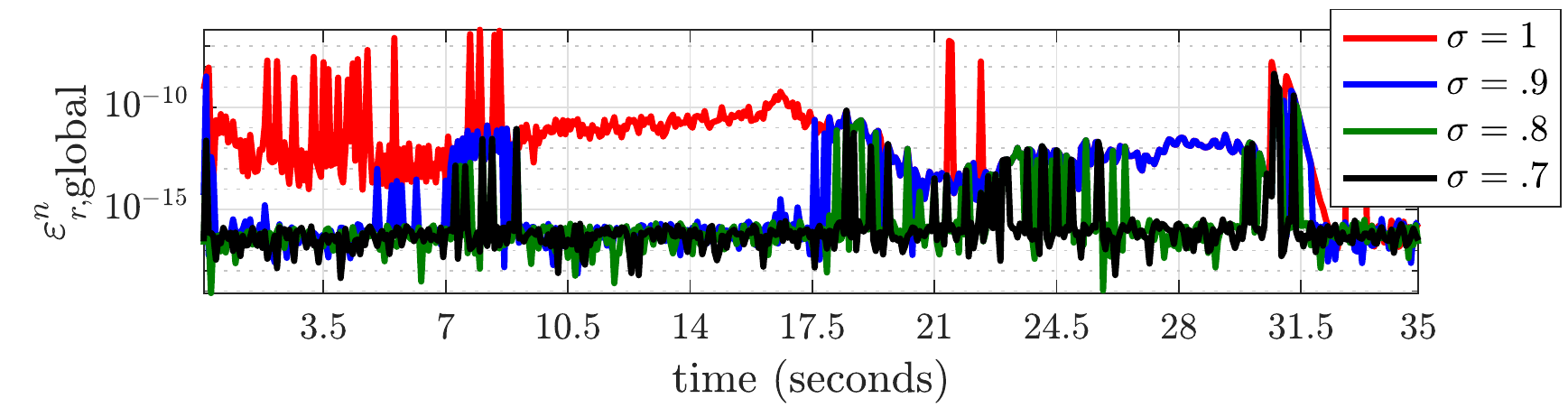}\label{fig:res_varying_sigma}}
    \caption{The time evolution of the time-instantaneous violation in global conservation
    $\resErrorNGlobal^n$ of \DCNameCap\ ($\nSubdomains=1$) for varying stopping tolerance $\tau$ and for varying scalar $\LagrangeScalar$. The autoencoder of the latent dimension $\nstatered=4$ is trained with the hybrid training objective function ($\rho=1$).}
	\label{fig:violation_instances_tol}
\end{figure}

Lastly, we explore different choices of parameters for SQP solvers
of \DCNameCap\ and their effects on the performance by measuring an additional error metric, time-instantaneous violation in global
conservation $\resErrorNGlobal^n \defeq \| \conserveOpGlobal \res^{n}(\aprxstate^{n}(\param);\param) \|_2^2$, $n=1,\ldots,\nseq$, which illustrates how violation in global conservation evolves in time.
First, Figure \ref{fig:res_varying_tol} reports the results of
\DCNameCap\ for varying stopping tolerance $\tau\in\{10^{-5}, 10^{-6},
10^{-7}, 10^{-8}\}$. The time-instantaneous violation in global conservation
$\resErrorNGlobal$ significantly decreases for stringent stopping tolerance,
while the averaged number of nonlinear iterations at each time step slightly
increases  $\{2.82, 2.94, 3.35, 3.73\}$. Second, we scale the update of the
Lagrange multipliers by a constant $\LagrangeScalar \in \{1, .9, .8, .7\}$ in the nonlinear iterations  (i.e., $\LagrangeScalar\delta \lagrangeMul^{n(k)}$). Figure \ref{fig:res_varying_sigma} shows that, as the scalar becomes smaller, the time-instantaneous violation in global conservation $\resErrorNGlobal$ decreases with the increased averaged number of nonlinear iterations $\{2.82, 3.42, 4.30, 5.17 \}$. From these observations, for achieving higher accuracy, simply increasing stopping tolerance for the original SQP solver (i.e., $\LagrangeScalar\!=\!1$) would be more desirable than using the SQP-variant with the additional scalar $\LagrangeScalar$. For the varying parameter values, there are only negligible changes in the state error and the error in the globally conserved variables.

\section{Conclusion}\label{sec:conc}
This work has proposed Deep Conservation: a novel  latent-dynamics learning technique that learns 
a nonlinear embedding using deep convolutional autoencoders, and computes a dynamics model via a projection process that enforces physical conservation laws. The dynamics model associates with a nonlinear least-squares problem with nonlinear equality constraints, and  the method requires the availability of a finite-volume discretization of the original dynamical system, which is used to define the objective function and constraints.
Numerical experiments  on an advection-dominated benchmark problem
demonstrated that Deep Conservation both achieves significantly higher
accuracy compared with classical projection-based methods, and guarantees the
time evolution of the latent state satisfies prescribed conservation laws. In
particular, the results highlight that \textit{both} the nonlinear embedding
\textit{and} the particular latent-dynamics model associating with the
solution to a constrained optimization problem are essential, as removing
either of these two elements yields a substantial degradation in performance.

\section{Acknowledgments}
This paper describes objective technical results
and analysis. Any subjective views or opinions that might be expressed in the paper do not necessarily
represent the views of the U.S. Department of Energy or the United States Government. Sandia
National Laboratories is a multimission laboratory managed and operated by National Technology
\& Engineering Solutions of Sandia, LLC, a wholly owned subsidiary of Honeywell International
Inc., for the U.S. Department of Energy’s National Nuclear Security Administration under contract
de-na0003525.


\appendix

\section{Latent-state trajectories}
\begin{figure}[!h]
    \centering
    \begin{minipage}{.475\linewidth}
    \subfloat[Latent trajectories of the training-parameter instances $\{(\paramelem{1}^{i}, \paramelem{2}^{0})\} = \{(4.25 + (1.25/9)i,\ 0.015)\}$, $i=0,\ldots,9$]  {\includegraphics[width=.9 \linewidth]{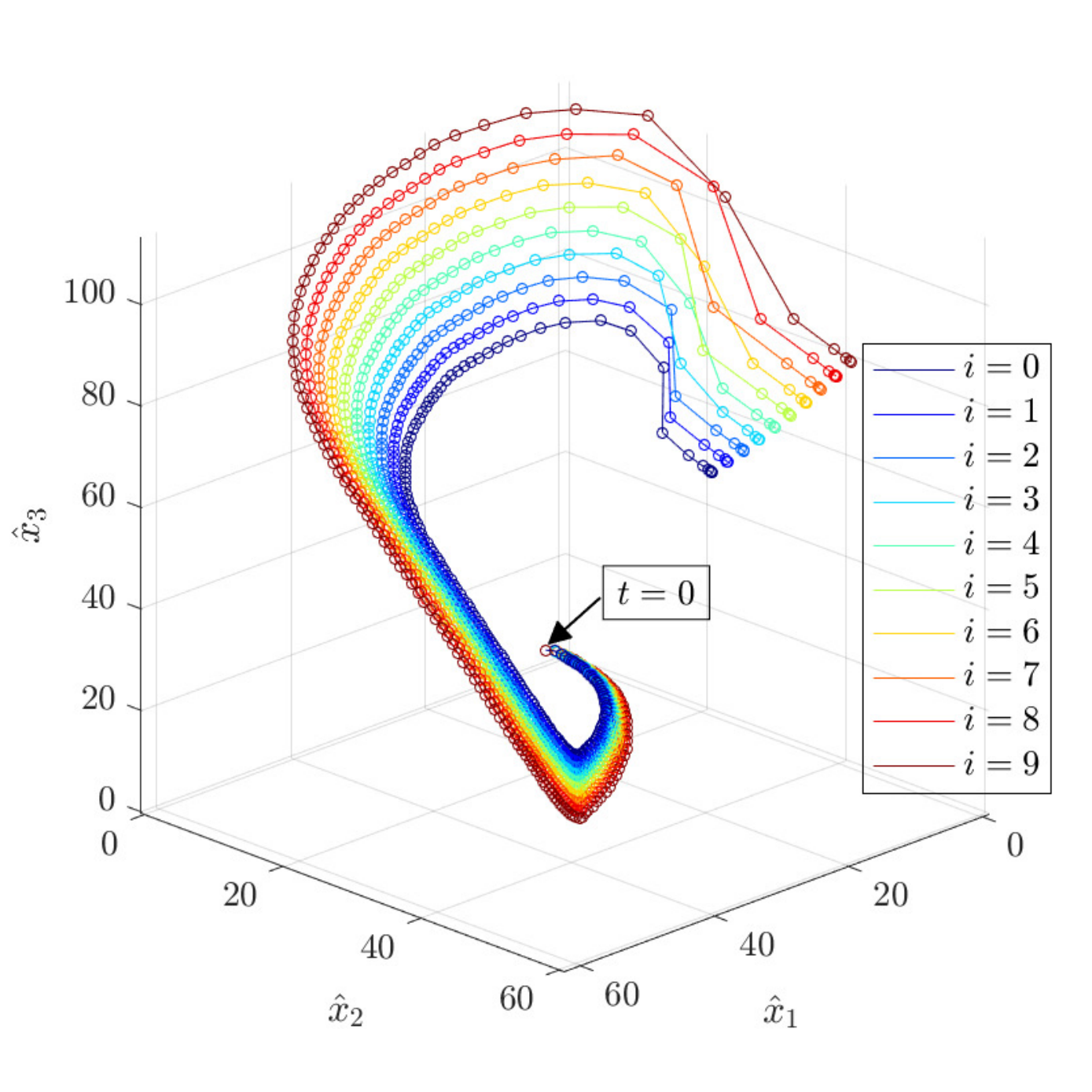}\label{fig:latent_state_traj_a} }
    \end{minipage}
    \begin{minipage}{.475\linewidth}
    \subfloat[Latent trajectories of the training-parameter instances $\{(\paramelem{1}^{0}, \paramelem{2}^{j})\} = \{(4.25, 0.015+(0.015/7)j)\}, j=0,\ldots,8$]  {\includegraphics[width=.9 \linewidth]{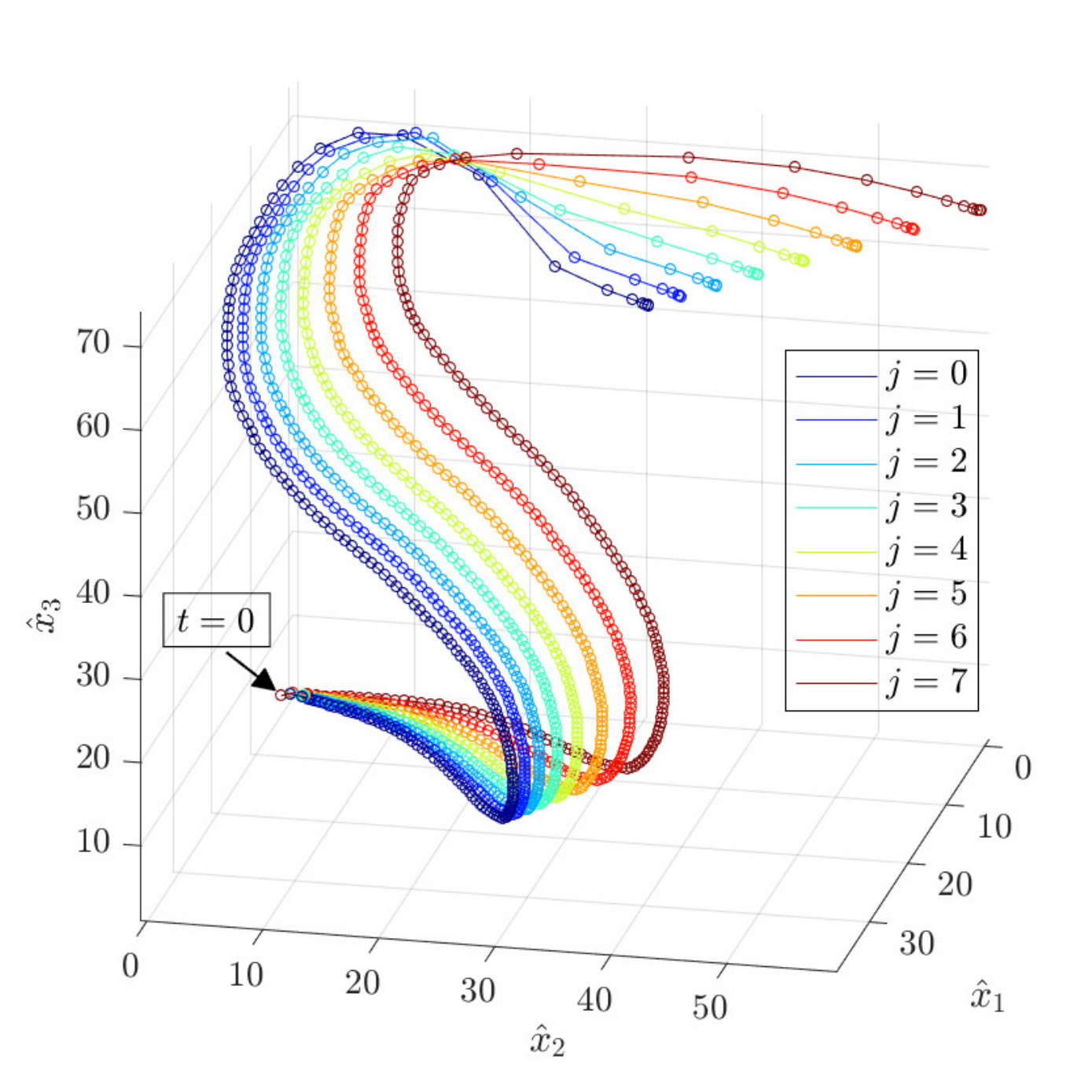}
    \label{fig:latent_state_traj_b}}
    \end{minipage}
    \caption{Example latent-state trajectories of the training data. The latent dimension of the autoencoder is $\dofrom=3$ and the latent states are obtained by applying the encoder to the collected FOM solution snapshots. The trajectories of latent states, which starts at $t=0$, for varying training-parameter instances are depicted, where the parameter domain is specified as $\paramspacetrain=\{(4.25 + (1.25/9)i,\ 0.015+(0.015/7)j)\}_{i=0,\ldots,9;\
j=0,\ldots 7}$.}
	\label{fig:latent_state_traj} 
\end{figure}

Figure \ref{fig:latent_state_traj} depicts trajectories of latent states $\rdstate^n(\paramtrain)$, $n=1,\ldots,\ntime$ of the training data (i.e., $\paramtrain \in \paramspacetrain$), where the dimension of the latent states is $\dofrom = 3$. A collection of latent states $\hat{\bm{\mathcal X}} \in \RR{\ntrain \times \dofrom \times \nseq} =  \RR{80 \times 3 \times 500}$ are obtained by applying the encoder $\encoder$ to the collected FOM solution snapshots $\bm{\mathcal X} \in \RR{\ntrain\times\nControlVol\times\nseq} =\RR{80 \times 512 \times 500}$, where the latent dimension of the autoencoder is $\dofrom=3$. Again, in our experiments, we choose 80 training-parameter instances on the uniform grid, $\paramspacetrain=\{(4.25 + (1.25/9)i,\ 0.015+(0.015/7)j)\}_{i=0,\ldots,9;\j=0,\ldots 7}$ and, for each training-parameter instance $\paramtrain^{\ell}$, we obtain a trajectory consisting of 500 latent states of dimension $3$.

Figure \ref{fig:latent_state_traj} depicts example latent trajectories. Figure \ref{fig:latent_state_traj_a} depicts the latent trajectories of the training-parameter instances $\{(\paramelem{1}^{i}, \paramelem{2}^{0})\} = \{(4.25 + (1.25/9)i,\ 0.015)\}$, $i=0,\ldots,9$, where the value of the first parameter is varying. And Figure \ref{fig:latent_state_traj_b} depicts the latent trajectories of the training-parameter instances $\{(\paramelem{1}^{0}, \paramelem{2}^{j})\} = \{(4.25, 0.015+(0.015/7)j)\}, j=0,\ldots,8$, where the value of the second parameter is varying. In both Figures, the latent trajectories start at the same point, indicated by $t=0$. This is because the initial condition of the inviscid Burgers' equation is the same for all the training-parameter instances (i.e., $\eqvar(\xoned,0;\param) = 1, \forall \xoned \in [0,100], \forall \param \in \paramspace$). From Figure \ref{fig:latent_state_traj} we observe that each training-parameter instance uniquely specifies the latent trajectory and its relative position in the latent space (i.e., latent trajectories of two parameter instances, that are close in the parameter space, are located closer than other latent trajectories).

\section{Hyperparameters}
For our experiments, we choose an autoencoder architecture, where the encoder consists of four convolutional layers followed by a fully-connected layer and the decoder consists of a fully-connected layer followed by four transposed-convolutional layers. With this network architecture, the first four convolutional layers extract feature maps of the solution snapshots on a coarse mesh (via strides larger than 1) and a fully-connected layer gathers the feature maps into a latent code. The decoder side performs the inverse of the encoder action. We keep the number of fully-connected layers in the decoder to one as we plan to further investigate adding more sparsity in the connections between each layer of the decoder in order to achieve higher computational efficiency and having consecutive fully-connected layers ruins the sparsity. 

We choose hyperparameters of this autoencoder architecture based on our observation that, for the given dataset and training strategy, increasing the network capacity (e.g., adding more (transposed) convolutional layers, add more filters, larger kernel size) does not achieve significantly better performance. On the other hand, decreasing the network capacity negatively affects the performance; we have explored the smaller number of kernel filters in the encoder and the decoder such as $\{\{8,16,32,64\}, \{32,16,8,1\}\}$, $\{\{4,8,16,32\}, \{16,8,4,1\}\}$, and $\{\{2,4,8,16\}, \{8,4,2,1\}\}$ or the smaller kernel filters such as $\{4,4,4,4\}$, which result in degradation of the performance. For the kernel filter sizes, in an effort to minimize producing checkerboard artifacts \cite{odena2016deconvolution}, we choose the kernel size that is divisible by the strides.

\bibliography{ref}

\begin{thebibliography}{10}

\bibitem{tensorflow2015-whitepaper}
{\sc M.~Abadi, A.~Agarwal, P.~Barham, E.~Brevdo, Z.~Chen, C.~Citro, G.~S.
  Corrado, A.~Davis, J.~Dean, M.~Devin, S.~Ghemawat, I.~Goodfellow, A.~Harp,
  G.~Irving, M.~Isard, Y.~Jia, R.~Jozefowicz, L.~Kaiser, M.~Kudlur,
  J.~Levenberg, D.~Man\'{e}, R.~Monga, S.~Moore, D.~Murray, C.~Olah,
  M.~Schuster, J.~Shlens, B.~Steiner, I.~Sutskever, K.~Talwar, P.~Tucker,
  V.~Vanhoucke, V.~Vasudevan, F.~Vi\'{e}gas, O.~Vinyals, P.~Warden,
  M.~Wattenberg, M.~Wicke, Y.~Yu, and X.~Zheng}, {\em {TensorFlow}: Large-scale
  machine learning on heterogeneous systems}, 2015.
\newblock Software available from tensorflow.org.

\bibitem{banijamali2017robust}
{\sc E.~Banijamali, R.~Shu, M.~Ghavamzadeh, H.~Bui, and A.~Ghodsi}, {\em Robust
  locally-linear controllable embedding}, in International Conference on
  Artificial Intelligence and Statistics, 2018.

\bibitem{benner2015survey}
{\sc P.~Benner, S.~Gugercin, and K.~Willcox}, {\em A survey of projection-based
  model reduction methods for parametric dynamical systems}, SIAM Review, 57
  (2015), pp.~483--531.

\bibitem{beucler2019achieving}
{\sc T.~Beucler, S.~Rasp, M.~Pritchard, and P.~Gentine}, {\em Achieving
  conservation of energy in neural network emulators for climate modeling},
  arXiv preprint arXiv:1906.06622,  (2019).

\bibitem{bohmer2015autonomous}
{\sc W.~B{\"o}hmer, J.~T. Springenberg, J.~Boedecker, M.~Riedmiller, and
  K.~Obermayer}, {\em Autonomous learning of state representations for control:
  An emerging field aims to autonomously learn state representations for
  reinforcement learning agents from their real-world sensor observations},
  KI-K{\"u}nstliche Intelligenz, 29 (2015), pp.~353--362.

\bibitem{carlbergGalDiscOpt}
{\sc K.~Carlberg, M.~Barone, and H.~Antil}, {\em Galerkin v.\ least-squares
  {P}etrov--{G}alerkin projection in nonlinear model reduction}, Journal of
  Computational Physics, 330 (2017), pp.~693--734.

\bibitem{carlberg2018conservative}
{\sc K.~Carlberg, Y.~Choi, and S.~Sargsyan}, {\em Conservative model reduction
  for finite-volume models}, Journal of Computational Physics, 371 (2018),
  pp.~280--314.

\bibitem{clevert2015fast}
{\sc D.~Clevert, T.~Unterthiner, and S.~Hochreiter}, {\em Fast and accurate
  deep network learning by exponential linear units {(ELUs)}}, in the 4th
  International Conference on Learning Representations, 2016.

\bibitem{cranmer2020lagrangian}
{\sc M.~Cranmer, S.~Greydanus, S.~Hoyer, P.~Battaglia, D.~Spergel, and S.~Ho},
  {\em Lagrangian neural networks}, arXiv preprint arXiv:2003.04630,  (2020).

\bibitem{demers1993non}
{\sc D.~DeMers and G.~W. Cottrell}, {\em Non-linear dimensionality reduction},
  in Advances in Neural Information Processing Systems, 1993, pp.~580--587.

\bibitem{fulton2019latent}
{\sc L.~Fulton, V.~Modi, D.~Duvenaud, D.~I. Levin, and A.~Jacobson}, {\em
  Latent-space dynamics for reduced deformable simulation}, in Computer
  graphics forum, vol.~38, Wiley Online Library, 2019, pp.~379--391.

\bibitem{glorot2010understanding}
{\sc X.~Glorot and Y.~Bengio}, {\em Understanding the difficulty of training
  deep feedforward neural networks}, in Proceedings of the Thirteenth
  International Conference on Artificial Intelligence and Statistics, 2010,
  pp.~249--256.

\bibitem{goodfellow2016deep}
{\sc I.~Goodfellow, Y.~Bengio, A.~Courville, and Y.~Bengio}, {\em Deep
  Learning}, vol.~1, MIT press Cambridge, 2016.

\bibitem{goroshin2015learning}
{\sc R.~Goroshin, M.~F. Mathieu, and Y.~LeCun}, {\em Learning to linearize
  under uncertainty}, in Advances in Neural Information Processing Systems,
  2015, pp.~1234--1242.

\bibitem{greydanus2019hamiltonian}
{\sc S.~Greydanus, M.~Dzamba, and J.~Yosinski}, {\em Hamiltonian neural
  networks}, in Advances in Neural Information Processing Systems, 2019,
  pp.~15353--15363.

\bibitem{hinton2006reducing}
{\sc G.~E. Hinton and R.~R. Salakhutdinov}, {\em Reducing the dimensionality of
  data with neural networks}, {Science}, 313 (2006), pp.~504--507.

\bibitem{hirsch2007numerical}
{\sc C.~Hirsch}, {\em Numerical Computation of Internal and External Flows: The
  Fundamentals of Computational Fluid Dynamics}, Elsevier, 2007.

\bibitem{holmes2012turbulence}
{\sc P.~Holmes, J.~L. Lumley, G.~Berkooz, and C.~W. Rowley}, {\em {Turbulence,
  Coherent Structures, Dynamical Systems and Symmetry}}, Cambridge University
  Press, 2012.

\bibitem{karl2016deep}
{\sc M.~Karl, M.~Soelch, J.~Bayer, and P.~van~der Smagt}, {\em Deep variational
  bayes filters: Unsupervised learning of state space models from raw data}, in
  International Conference on Learning Representations, 2017.

\bibitem{kingma2014adam}
{\sc D.~P. Kingma and J.~Ba}, {\em {Adam: A method for stochastic
  optimization}}, in the 3rd International Conference on Learning
  Representations, 2015.

\bibitem{lecun2015deep}
{\sc Y.~LeCun, Y.~Bengio, and G.~Hinton}, {\em Deep learning}, Nature, 521
  (2015), p.~436.

\bibitem{lee2020model}
{\sc K.~Lee and K.~T. Carlberg}, {\em Model reduction of dynamical systems on
  nonlinear manifolds using deep convolutional autoencoders}, Journal of
  Computational Physics, 404 (2020), p.~108973.

\bibitem{lesort2018state}
{\sc T.~Lesort, N.~D{\'\i}az-Rodr{\'\i}guez, J.-F. Goudou, and D.~Filliat},
  {\em State representation learning for control: An overview}, Neural
  Networks,  (2018).

\bibitem{leveque2002finite}
{\sc R.~J. LeVeque}, {\em Finite volume methods for hyperbolic problems},
  vol.~31, Cambridge university press, 2002.

\bibitem{lusch2018deep}
{\sc B.~Lusch, J.~N. Kutz, and S.~L. Brunton}, {\em Deep learning for universal
  linear embeddings of nonlinear dynamics}, Nature communications, 9 (2018),
  p.~4950.

\bibitem{mojgani2017lagrangian}
{\sc R.~Mojgani and M.~Balajewicz}, {\em Lagrangian basis method for
  dimensionality reduction of convection dominated nonlinear flows}, arXiv
  preprint arXiv:1701.04343,  (2017).

\bibitem{morton2018deep}
{\sc J.~Morton, A.~Jameson, M.~J. Kochenderfer, and F.~Witherden}, {\em Deep
  dynamical modeling and control of unsteady fluid flows}, in Advances in
  Neural Information Processing Systems, 2018, pp.~9258--9268.

\bibitem{odena2016deconvolution}
{\sc A.~Odena, V.~Dumoulin, and C.~Olah}, {\em Deconvolution and checkerboard
  artifacts}, Distill, 1 (2016), p.~e3.

\bibitem{ohlberger2016reduced}
{\sc M.~Ohlberger and S.~Rave}, {\em Reduced basis methods: Success,
  limitations and future challenges}, in Proceedings of ALGORITMY, Slovak
  University of Technology, 2016, pp.~1--12.

\bibitem{otto2019linearly}
{\sc S.~E. Otto and C.~W. Rowley}, {\em Linearly recurrent autoencoder networks
  for learning dynamics}, SIAM Journal on Applied Dynamical Systems, 18 (2019),
  pp.~558--593.

\bibitem{raissi2019physics}
{\sc M.~Raissi, P.~Perdikaris, and G.~E. Karniadakis}, {\em Physics-informed
  neural networks: A deep learning framework for solving forward and inverse
  problems involving nonlinear partial differential equations}, Journal of
  Computational Physics, 378 (2019), pp.~686--707.

\bibitem{takeishi2017learning}
{\sc N.~Takeishi, Y.~Kawahara, and T.~Yairi}, {\em Learning {K}oopman invariant
  subspaces for dynamic mode decomposition}, in Advances in Neural Information
  Processing Systems, 2017, pp.~1130--1140.

\bibitem{toth2019hamiltonian}
{\sc P.~Toth, D.~J. Rezende, A.~Jaegle, S.~Racani{\`e}re, A.~Botev, and
  I.~Higgins}, {\em Hamiltonian generative networks}, arXiv preprint
  arXiv:1909.13789,  (2019).

\bibitem{watter2015embed}
{\sc M.~Watter, J.~Springenberg, J.~Boedecker, and M.~Riedmiller}, {\em {Embed
  to control: A locally linear latent dynamics model for control from raw
  images}}, in Advances in Neural Information Processing Systems, 2015,
  pp.~2746--2754.

\bibitem{wiewel2019latent}
{\sc S.~Wiewel, M.~Becher, and N.~Thuerey}, {\em Latent space physics:
  {T}owards learning the temporal evolution of fluid flow}, in Computer
  Graphics Forum, vol.~38, Wiley Online Library, 2019, pp.~71--82.

\bibitem{zahr2015progressive}
{\sc M.~J. Zahr and C.~Farhat}, {\em Progressive construction of a parametric
  reduced-order model for {PDE}-constrained optimization}, International
  Journal for Numerical Methods in Engineering, 102 (2015), pp.~1111--1135.

\end{thebibliography}
\bibliographystyle{siam}

\end{document}